\documentclass[
 reprint, prl,
onecolumn,
 amsmath,amssymb,
 aps,
]{revtex4-2}

\usepackage{graphicx}
\usepackage{dcolumn}
\usepackage{bm}
\usepackage{tikz}
\usepackage{comment}
\usepackage{amssymb}
\usepackage{navigator}

\usepackage{amsfonts,amsmath,amssymb,amsthm}
\usepackage{xcolor,colortbl}

\theoremstyle{plain}

\usepackage{color}

\def\tmin{T_{\text{min}}}
\def\tmax{T_{\text{max}}}

\begin{document}

\preprint{APS/123-QED}

\title{Noise-induced transition to stop-and-go waves in single-file traffic \\ rationalized by an analogy with Kapitza's inverted pendulum}

\author{Oscar Dufour}
\author{Alexandre Nicolas}
\author{David Rodney}
\affiliation{
Université Claude Bernard Lyon 1, CNRS, Institut Lumière Matière, UMR 5306, F-69100, Villeurbanne, France
}

\author{Jakob Cordes}
\altaffiliation{Also at Institute of Advanced Simulation, Forschungszentrum Jülich GmbH, Jülich, Deutschland}
\author{Andreas Schadschneider}
\altaffiliation{Also at Institut für Physikdidaktik, Universität zu Köln, Köln, Deutschland}
\affiliation{
 Institut für Theoretische Physik, Universität zu Köln, Köln, Deutschland
}

\author{Antoine Tordeux}
\affiliation{
 Fakultät für Maschinenbau und Sicherheitstechnik, Bergische Universität Wuppertal, Wuppertal, Deutschland
}

\newcommand{\comm}[1]{}
\newcommand{\bs}[1]{\boldsymbol{#1}}
\date{\today}

\begin{abstract}
Stop-and-go waves in vehicular traffic are commonly explained as a linear collective instability induced by e.g. response delays. We explore an alternative mechanism that more faithfully mirrors oscillation formation in dense single-file traffic. Stochastic noise plays a key role in this model; as it is increased, the base (uniform) flow abruptly switches to stop-and-go dynamics despite its unconditional linear stability. We elucidate the instability mechanism and rationalize it quantitatively by likening the system to a cyclically driven Kapitza pendulum.

\bigskip

 
\end{abstract}

\maketitle

Analyzing the stability of many-body systems ($N\geqslant 3$) is often a prickly issue, as illustrated by the longstanding debate over the stability of the solar system. Further complications arise when the system's dynamics cannot be described by deterministic equations of motion \cite{gardiner2009stochastic}. 
Vehicular traffic perfectly illustrates this difficulty.  
Although every driver is familiar with stop-and-go waves on highways (where traffic jams, sometimes called phantom jams, 
emerge for no apparent reason, forcing drivers to slow down and speed up again),
the underlying physical origin is still a matter of debate, even for single-file traffic.
This is exemplified by a recent suggestion that, in the context of three-phase traffic theory \cite{Kerner23}, traffic breakdown could be caused by the over-acceleration of drivers \cite{Kerner23}.
The debate has practical implications for safety and congestion, recently re-ignited by the instabilities observed in platooning experiments with vehicles equipped with adaptive cruise control (ACC) systems \cite{stern_DissipationStopandgoWaves_2018,gunter2020commercially,makridis2021openacc}. 
A plethora of factors can destabilize single-file traffic flow, as vehicle control and environmental perception are subject to reaction times, delays \cite{nagatani1998delay}, and inaccuracies in perception or response \cite{yeo2009understanding,laval2014parsimonious}. Several of these factors can be handled deterministically.
In particular, finite response times and latency in vehicle control \cite{nagatani1998delay,orosz2004global,orosz2010traffic,wilson_CarfollowingModelsFifty_2011,tordeux_LinearStabilityAnalysis_2012,tordeux2018traffic} can be modeled by inertial ordinary differential equations \cite{komatsu1995kink} and delayed linear equations \cite{nagatani1998delay}. Both theoretical  \cite{reuschel1950fahrzeugbewegungen,pipes_OperationalAnalysisTraffic_1953,kometani1958stability,chandler1958traffic,herman_TrafficDynamicsAnalysis_1959,komatsu1995kink,orosz2010traffic} and numerical studies \cite{bando_DynamicalModelTraffic_1995,bando1998analysis} have traditionally identified the foregoing factors as causes of (linear) instabilities that drive the platoon or string of vehicles away from uniform steady flow.
The stop-and-go dynamics that ensue when the response becomes too slow \cite{makridis2019response} reproduce the amplification of spontaneous perturbations observed in experiments with ACC-equipped vehicles \cite{gunter2020commercially,makridis2021openacc}, notably in fine-tuned nonlinear models \cite{treiber_CongestedTrafficStates_2000,jiang_FullVelocityDifference_2001,tomer2000presence,tordeux2010adaptive}.

However, for cars without ACC, these delay-induced instabilities arguably fail to capture prominent features observed in naturalistic single-file traffic \cite{jiang2018experimental,tian2016empirical} as well as  in controlled experiments with platoons of vehicles  \cite{jiang2015some,jiang2018experimental,tian2019role} or cars on a ring \cite{sugiyama2008traffic,tadaki2013phase}. In the latter experiments, around twenty drivers followed each other along a large single-file loop. 
Above a critical density, stop-and-go waves emerged. However, rather than arising from the gradual amplification of an initial perturbation, they appeared after stochastic incubation periods ranging from one to several minutes
\cite{sugiyama2008traffic,stern_DissipationStopandgoWaves_2018}. Such irregular emergence of stop-and-go waves is incompatible with linear instability. Instead, it points to a metastable (and \emph{not} linearly unstable) state \cite{nakayama2009metastability}  \cite[Chap. 6.3-5]{schadschneider_StochasticTransportComplex_2010}. 
Moreover, linearly unstable systems that directly relate speed and spacing cannot account for the concave growth of speed oscillations observed along a platoon of cars following a leader driving at constant speed \cite{treiber2017intelligent,jiang2018experimental,tian2019role}.
In contrast, concave growth can be almost quantitatively reproduced by car-following models featuring (deterministic) action points, i.e., in which a response is triggered only if the stimulus exceeds a finite threshold \cite{treiber2017intelligent}.

Alternatively, this concave growth of fluctuations is equally well reproduced by complementing the equation of motion with a stochastic term \cite{treiber2017intelligent,jiang2018experimental}, viz
\begin{equation}
\mathrm{d} v_n(t)=F\big(\Delta x_n(t),v_n(t),v_{n+1}(t)\big)\,\mathrm{d}t + \sigma\mathrm{d} W_n(t)\label{eq:basic}.
\end{equation}
The stochastic noise $\sigma\mathrm{d} W_n$ describes the combined effect of human error and of the many degrees of freedom that have been left out of the deterministic response  $F$, which here depends on the gap $\Delta x_n=x_{n+1}-x_n-\ell$ ($x_n$ is the position of the $n$-th car, $x_{n+1}$ the position of the predecessor, i.e., the $n$+1-th car, and $\ell\ge0$ the car length), the speed $v_n=\dot x_n$, and the speed of the predecessor $v_{n+1}(t)$. For systems exhibiting linear instability in some range of parameters, 
a pronounced effect of the noise is expected.
Indeed, even in strictly physical systems \cite{wiesenfeld1985noisy}, `noisy precursors' arise near dynamical instabilities in the form of e.g. sustained oscillations. These oscillations are also seen in \emph{stable} linear car-following models, in which noise is added to account for human error or idiosyncrasies \cite{laval2014parsimonious,jiang2018experimental}. But they should not be mistaken for a genuine instability: it is known that \emph{additive} noise cannot modify the stability of a linearized stochastic differential equation 
\cite{gardiner2009stochastic}. Instead of additive noise, to uncover a \emph{bona fide} instability,  a \emph{multiplicative} noise was employed in Ref.~\cite{Ngoduy21}, which grows with the relative speeds. Schematically, the noise is 
amplified as 
the system gradually departs from uniform steady flow so that, starting from a none-too-stable state, this positive feedback mechanism can push the system into instability. 
Crucially, in these models, the noise acts \emph{perturbatively}: stability or instability of the system depends on the properties of the deterministic response function $F$ around the stationary flow \cite{treiber2017intelligent,Ngoduy21}, and the noise merely amplifies or `anticipates' the growth of deterministically unstable modes.

In this Letter, we challenge the prevailing view that stop-and-go waves arise from a deterministic instability of the stationary flow, amplified by noise. We show that even a simple additive noise can
fully destabilize a car-following model which is unconditionally linearly stable in the deterministic limit.
We elucidate the nonlinear instability mechanism by drawing on an analogy with the Kapitza pendulum, 
revealing its inherently nonperturbative nature.
This finding calls into question the widespread tendency in the literature to focus primarily on the \emph{local} properties of the deterministic response function $F$.


Traffic problems have given rise to models galore. 
Insight into metastable states was in part gained using 
cellular automata and interacting particle systems \cite{BarlovicSSS98,Ke-Ping_2004,kaupuvzs2005zero,huang2018instability}, \cite[Chap. 8.1]{schadschneider_StochasticTransportComplex_2010}. 
Instead, here we target continuous (in time and space) car-following models of the generic structure of Eq.~\ref{eq:basic}
that can be either deterministic ($\mathrm{d}W_n=0$) or complemented with a stochastic term. 
We have studied a number of models in this broad class, including the stochastic full velocity difference (SFVD) model
\cite{wagner2011time,treiber2009hamilton,wang2020stability,friesen2021spontaneous}, the inertial car-following model of Tomer et al. \cite{tomer2000presence}, and the stochastic intelligent driver (SID) model \cite{treiber2017intelligent}, detailed in the \emph{End Matter}.
Our focus will be put here on the stochastic adaptive time gap (SATG) model, given by 
\begin{equation}
    \mathrm{d}v_n =\frac{1}{T_n^\varepsilon}\big(\lambda \left(\Delta x_n-Tv_n\right)+\Delta v_n\big)\,\mathrm{d}t+\sigma \mathrm{d}W_n,
    \label{SATG}
\end{equation}
where $\lambda>0$ is a sensitivity parameter that governs how intent drivers are on keeping a desired time gap $T$ with their predecessor, and $T_n^\varepsilon$ a bounded molifier of the time gap. 
These models are used to simulate a single file of cars on a circuit (with periodic boundary conditions and no overtaking) to mimic the experimental settings of Refs.~\cite{sugiyama2008traffic,tadaki2013phase} (see \emph{End Matter} for details).

\begin{figure}[!ht]
    \centering
    \includegraphics[width=.47\columnwidth]{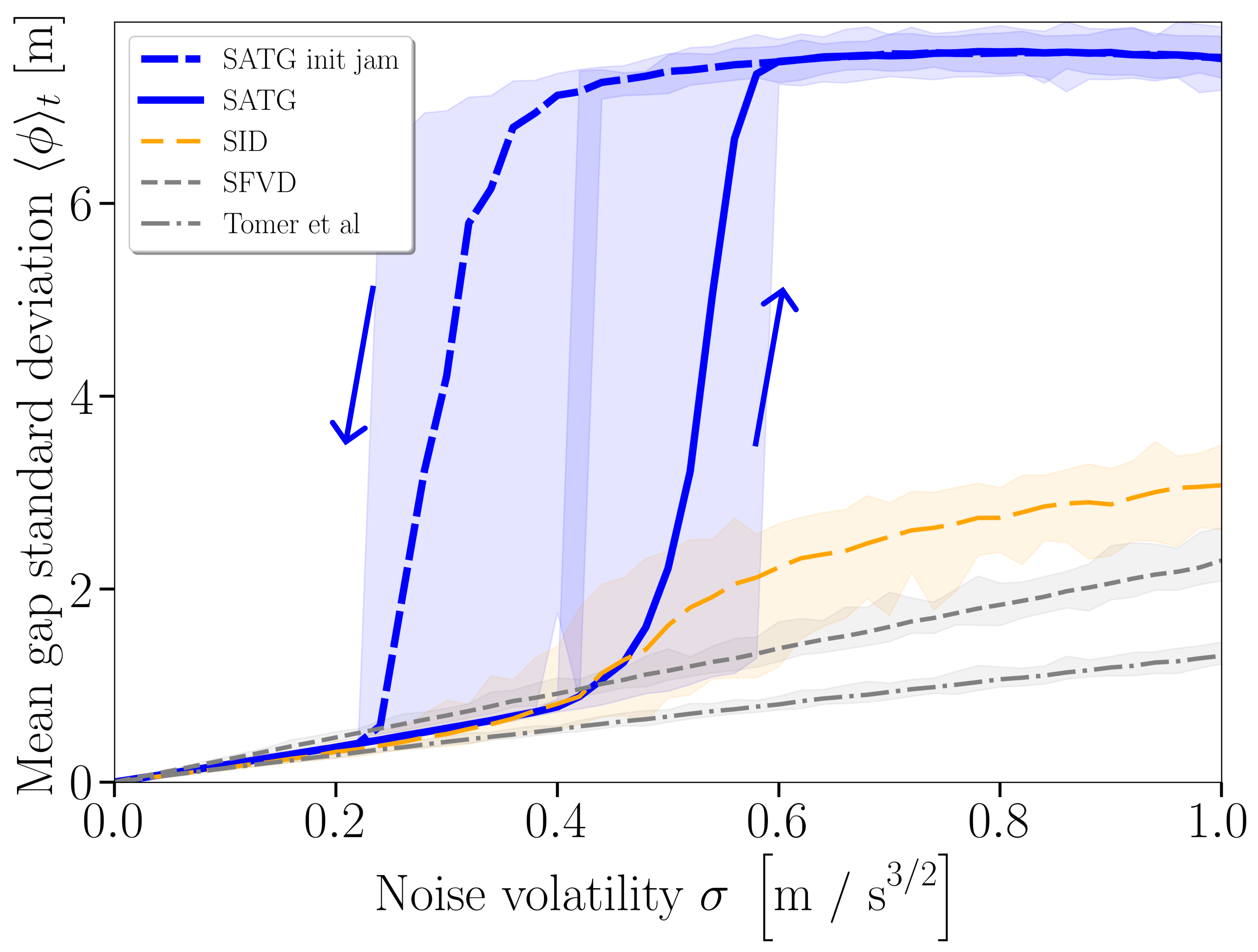}\vspace{-3mm}
    \caption{Mean standard deviation $\langle \phi \rangle_t$ (where $\langle \cdot \rangle_t$ denotes a time average) of the gaps in stationary state as a function of the noise volatility $\sigma$, for $N=22$ cars in the experimental settings of \cite{sugiyama2008traffic} (see \emph{End Matter} for details). The colored overlays show the min/max range. Note that the near-linear (SFVD and Tomer et al.) models show no noise-induced instability, whereas the nonlinear SID and SATG models undergo a transition to stop-and-go dynamics. The dashed blue line shows SATG simulations starting from a jammed initial condition, pointing to significant hysteresis.}
    \label{fig1}
\end{figure}

Linear stability analysis predicts that the regime of uniform, time-independent flow is stable at low car densities (for vanishing noise $\sigma\to 0$) but gives way to an instability (resulting in stop-and-go waves) at higher density in almost all models, at least for some range of parameters \cite{orosz2004global,treiber2017intelligent}.
The SATG model stands out in this respect, being unconditionally linearly stable around the uniform flow state (see Sec.~B of the Supplementary Information and Refs.~\cite{khound2021extending, ehrhardt2024stability}).
Direct numerical simulations of the models for $N=22$ cars confirm the expected stability for low noise $\sigma$.  This is shown in Fig.~\ref{fig1} using the disorder parameter $\phi(t) = \sqrt{ \overline{ \Delta x_n^2}(t)  -  \overline{ \Delta x_n }^2(t)}$, where the overline denotes an average over the $N$ cars. The gradual increase of the time average $\langle \phi \rangle_t$ with $\sigma$ in the SFVD and Tomer models reflects the increasing fluctuations in inter-vehicular gaps induced by noise. In the SID model, disorder rises more markedly (but still fairly smoothly) around $\sigma \approx 0.5\,\mathrm{m/s^{3/2}}$, mirroring the excitation of none-too-stable modes (here, stop-and-go waves) in the vicinity of a deterministic instability. 
In contrast,  in the SATG model, the simulations reveal an unexpected, abrupt transition to traffic oscillations (high $\phi$ values) when $\sigma$ crosses a threshold $\sigma^{\star}$.
For $\sigma>\sigma^{\star}$ (where $\sigma^{\star}$ decreases with increasing system size), 
once developed, the stop-and-go waves are similar in shape to those found in other models, with a wavelength given by the system size, but more pronounced. The jammed phase propagates upstream, at speed $v_c=-\ell/T\approx -5$~m/s with $\ell$ the vehicle length and $T$ the desired time gap (in SATG). Besides, since the noise acts on the acceleration and remains moderate, the SATG trajectories remain smooth.
 These features are in line with experimental observations (see Fig.~\ref{fig:traj}  in the \emph{End Matter}; also see Fig.~{\bf S1} for fine-tuned model parameters). 


In addition to its abruptness, the SATG route to instability stands apart in that the emergence of stop-and-go waves occurs erratically after a typically fairly long but broadly varying transient time (Fig.~\ref{fig:TTJ} in the \emph{End Matter}), unlike the waves originating from a linear instability. Such a transient is consistent with the empirical data of Ref.~\cite{sugiyama2008traffic,tadaki2013phase}, where stop-and-go waves do not occur systematically for a given car density.  It points to a first-order transition towards traffic instability and to metastability of both the jammed and unjammed states, in accordance with the empirical conclusions of Ref.~\cite{nakayama2009metastability} and the numerical finding of a bimodal distribution of $\phi$ at the transition, in Fig.~{\bf S2}. 

As a corollary of metastability, we expect hysteresis -- a common observation in highway traffic 
\cite{chen2012microscopic,yeo2009understanding}. Our simulations confirm that a significant hysteresis loop arises when the system starts from the stop-and-go regime and the noise volatility (or the car density) is gradually reduced, as shown in Fig.~\ref{fig1}: stop-and-go waves subsist below the noise level required to destabilize the homogeneous flow. However, when noise further 
decreases, the system resumes uniform flow before reaching the deterministic limit $\sigma=0$. This implies that noise does not simply trigger the instability, as in a subcritical hydrodynamic instability where the system leaps into another flow branch, but is required to sustain the stop-and-go dynamics, in the same way as temperature is needed to stabilize the gas phase past the boiling transition.
Furthermore, unlike the SID case \cite{treiber2017intelligent}, the destabilization of the uniform flow cannot  be ascribed to poorly damped perturbations in the relative vicinity of a linear instability (for want of being able to identify such an instability), nor to the self-reinforcement of a multiplicative noise, as in \cite{Ngoduy21}.
Lastly, in the absence of a bifurcation in the deterministic limit, the noise-induced SATG instability does not lend itself to continuation methods \cite{orosz_GlobalBifurcationInvestigation_2004}, nor to perturbative stochastic stability analysis \cite{gardiner2009stochastic}.

\vspace{0.5cm}

We hypothesize that the (finite) noise operates as an external driving force whose continuous actions destabilize the (deterministic) fixed point and stabilize stop-and-go dynamics. Concretely, we draw an analogy with a mechanical system: The Kapitza pendulum is a rigid pendulum subjected to vertical oscillatory vibrations which may stabilize its `inverted' equilibrium position, here amalgamated to the state with stop-and-go waves. Thus, we split the noise into a controlled part (periodic driving) and an uncontrolled part $\mathrm{d}\widehat{W}$ (the residual stochastic perturbation).
More precisely, considering a realization of the noise $\bigl(\sigma \mathrm{d}W_n\bigr)$ over a large time interval,
we arbitrarily isolate one of its Fourier modes and inject  it into the deterministic Eq.~\eqref{SATG} as a (controlled) oscillatory driving 
$\bigl\{ C \cos(\omega t + \varphi_n) \bigr\}$, with $C>0$ and random phases $\varphi_n$ \footnote{The idea of introducing oscillations also pops up when one writes an amplitude equation in hydrodynamics \cite{morozov2005subcritical}, but in that case it helps probe the nonlinear growth of arbitrary perturbations.}; the residual signal is handled as perturbative white noise $\widehat{\sigma}\,\mathrm{d} \widehat{W}_n$. Thus, Eq.~\eqref{SATG} turns into
\begin{equation}
\mathrm{d} v_n(t)=F_n\big(t)\,\mathrm{d}t + C\,\cos(\omega t +\varphi_n) \,\mathrm{d}t + \widehat{\sigma}\,\mathrm{d} \widehat{W}_n(t)\label{eq:cyclic_basic},
\end{equation}
where $F_n(t)=:F\big(\Delta x_n(t),v_n(t),v_{n+1}(t)\big)$.

\begin{figure*}[!ht]
    \centering
    \includegraphics[width=.75\textwidth]{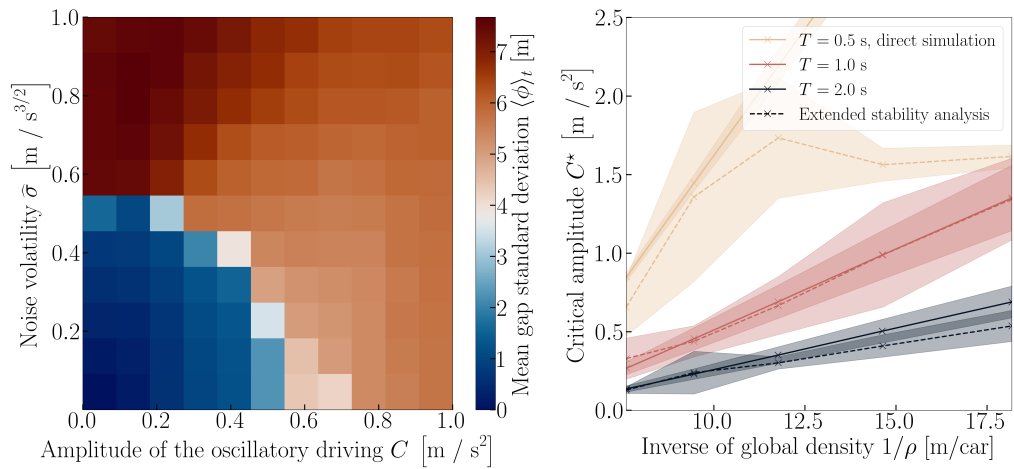}
    \caption{Left panel: Stability diagram in which the noise effect is heuristically split into a controlled oscillatory driving of amplitude $C$ and a residual random noise $\widehat{\sigma}$. Color overlay: standard deviation of the gaps. Right panel: Variation of the stability threshold $C^{\star}$ with the car density, computed with direct simulations of Eq.~\eqref{eq:cyclic_basic} at a driving frequency of $0.05\cdot 2\pi\,\mathrm{rad/s}$ or derived from an extended stability analysis (see main text).}
    \label{fig:StabDia}
\end{figure*}

The numerical simulations presented in Fig.~\ref{fig:StabDia}, left panel, confirm that oscillatory driving facilitates the emergence of stop-and-go waves: the larger the driving amplitude $C$, the smaller the perturbative noise to trigger the instability. For strong enough driving $C\geqslant C^{\star}$ (with $C^{\star}=0.55\,\mathrm{m/s^2}$ on Fig.~\ref{fig:StabDia}), the system becomes  unstable under vanishing noise. Since  the noise in Eq.~\eqref{SATG} contains a density $\sigma$ of  the isolated Fourier mode, we hypothesize that $C^{\star}$ and $\sigma^{\star}$ are linearly related, i.e.,
$C^{\star}=A(\lambda,T)\,\sigma^{\star}$, with a coefficient $A(\lambda,T)$ of order 
$1\,\mathrm{s^{-1/2}}$ that may depend on model parameters $\lambda$ and $T$. This is indeed the case: $C^{\star}\simeq 1.0\,\sigma^{\star}$ holds when the density is varied, for all angular frequencies $\omega$ up to  $ 0.05\cdot 2\pi\,\mathrm{rad/s}$. In Appendix ({\bf D}), we derive a theoretical expression for $A(\lambda,T)$, by equating the mean-square gap fluctuations $\langle |\Delta x_n|^2 \rangle$ expected for white noise and for oscillatory driving; the expression agrees with the observations only semi-quantitatively, but it satisfactorily captures the insensitivity of $A(\lambda,T)$ to car density and frequency, and its variations with model parameters. 
The foregoing parallel elucidates the mechanism by which the instability unfolds  in the genuine SATG model: finite noise continuously excites nonlinearities in the deterministic response function $F$, as does the oscillatory driving, which brings the system to the brink of stability. Accordingly, in contrast with other mechanisms, the instability does not result from \emph{local} properties of $F$ around the state of uniform flow but from \emph{nonlocal} properties.

The oscillatory driving opens the door to a more quantitative rationalization. At $\omega=0$, the driving is equivalent to acceleration offsets $b_n=C\,\cos(\varphi_n)$ applied to each car. It turns out that these heterogeneous offsets can make the system linearly unstable for a range of offsets $b_n$, and implicit analytical expressions can be derived for the growth rates, as we show in Sec.~C of the SI. Let $\nu(\{b_n\})$ be the largest nontrivial growth rate (i.e. the real part of the eigenvalue).
If the oscillatory frequency $\omega$ is small but nonzero, the driving can be treated as quasi-stationary. Under this approximation, at each time $t'$, the peak perturbation grows exponentially as $\exp(\nu_{t'} t)$, where $\nu_{t'} = \nu(\{C \cos(\omega t' + \varphi_n)\})$.
To go further, since the peak growth mode mostly spans the system size, we overlook the fact that it may change with time $t'$, so that the effective growth rate over a cycle is
 \begin{eqnarray}
     \nu_{\mathrm{eff}} (C) &=& \langle\nu_{t'} \rangle_{t'}  \nonumber \\
     & \simeq  & \langle \nu(\{C\, \cos(\theta_n)\}) \rangle_{\theta},
     \label{eq:lambda}
 \end{eqnarray}
 where the angular brackets denote averages over time $t'$ or over uniformly distributed $\{\theta_n\}$. In the last approximation, we replaced the actual phases $\omega t'+\varphi_n$  with random ones $\theta_n$ drawn from a uniform distribution over $]-\pi,\pi]$. This yields a very good agreement after averaging over the random phases ${\varphi_n}$, as observed numerically in Fig.~{\bf S3} of the SI. 
 The thresholds $C^{\star}$ at which $\nu_{\mathrm{eff}} (C)$ becomes positive are determined by numerically computing the eigenvalues of Eq.~\ref{eq:lambda}.
 Remarkably, Fig.\ref{fig:StabDia}, right panel, proves that the values of $C^{\star}$ predicted by the present extended stability analysis accurately reproduce the instability thresholds measured in direct simulations of Eq.~\eqref{eq:cyclic_basic} under oscillatory driving, even at finite frequencies (deviations are observed for small $T$ at low density). It follows that these calculations succeed in inferring the noise threshold $\sigma^{\star}=
 C^{\star} / A$, with $A=1.0\,\mathrm{s^{-1/2}}$, at which the genuine SATG model (Eq.~\ref{SATG}) undergoes a transition (Fig.~\ref{fig1}). 

\bigskip

\vspace{0.5cm}

\begin{figure*}[!ht]
    \centering\vspace{-2mm}
    \includegraphics[width=.8\textwidth]{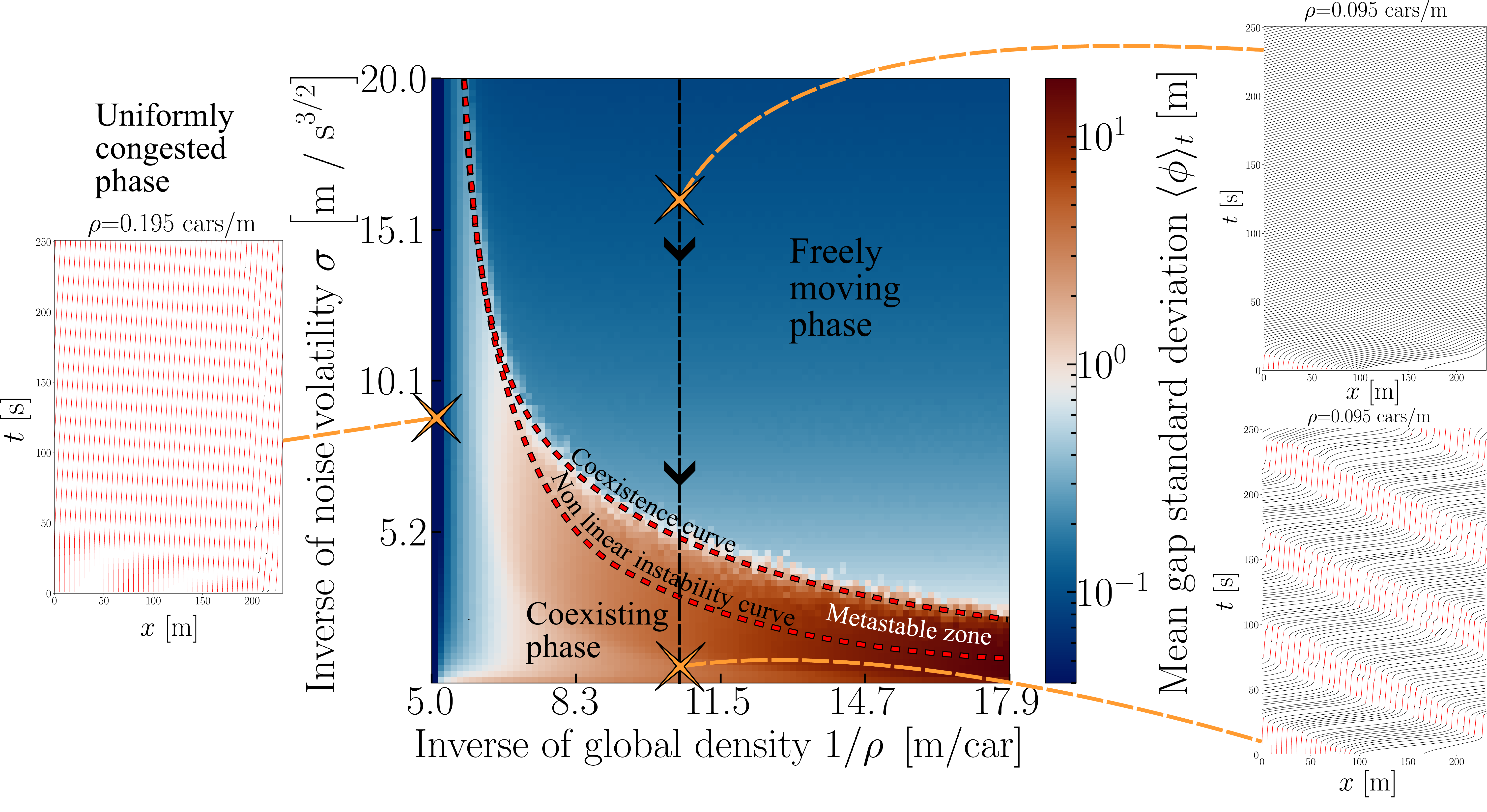}
    \caption{\emph{Main panel:} `Phase' diagram of SATG traffic as a function of the inverse car density (i.e., the distance between the mid-points of cars) and the inverse noise volatility; the heat map represents the gap standard deviation $\langle \phi \rangle$. The side panels show kymographs of trajectories in different regions of parameter space.}
    \label{fig:PhaseDiag}
\end{figure*}

\begin{figure}[!ht]
    \centering
    \includegraphics[width=.75\columnwidth]{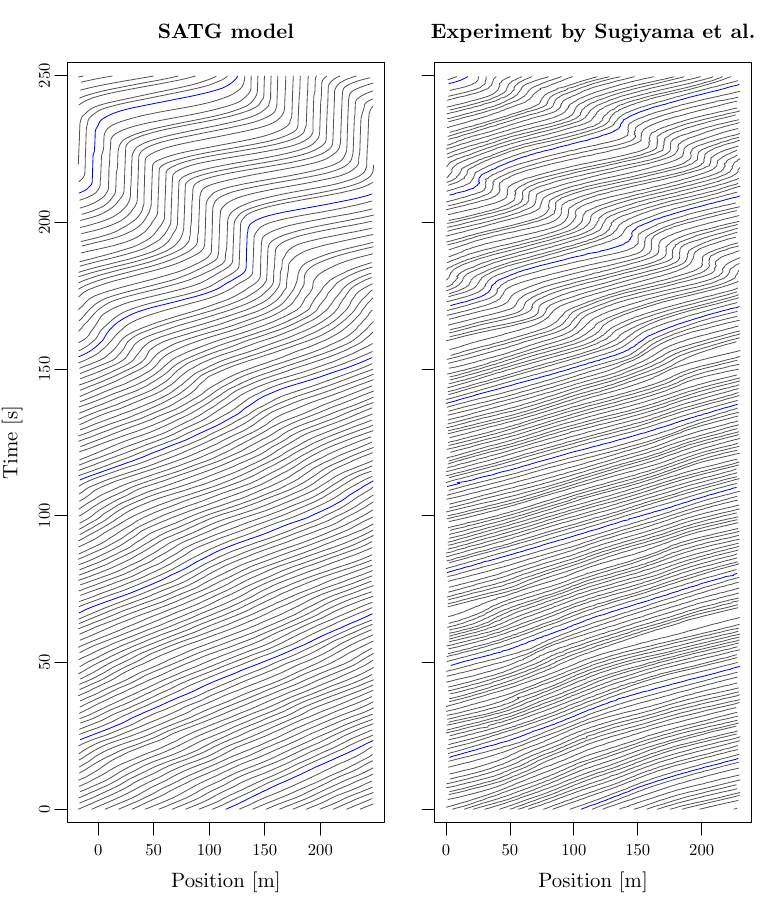}
    \caption{Trajectories, featuring stop-and-go waves for a selected simulation of the SATG model \protect\eqref{SATGapp}-\protect\eqref{SATGapp2} with $\sigma=0.6$~m/s$^{3/2}$ (left panel), and in the experiment of Sugiyama et al. \protect\cite{sugiyama2008traffic} (right panel).}
    \label{fig:traj}
\end{figure}

The above discussion
bolsters the relevance of the Kapitza analogy. However, despite the singularity of this destabilizing mechanism, from a broader perspective, we find that the general picture of a first-order phase transition established for traffic instabilities \cite{KernerR97,nagatani2002physics} originating from other processes, e.g., reaction delay \cite{nagatani1998thermodynamic}, slow-to-start effects \cite{BarlovicSSS98} or cellular automata simulated in continuous space \cite{jost2005probabilistic}, still holds. 
To clarify the picture, let us map the transition to traffic oscillations onto a liquid-gas transition \cite{NagelWW03} by assimilating the noise volatility $\sigma$ to the temperature and the inverse headway (local density $\rho$) between cars to the density.
In Fig.~\ref{fig:PhaseDiag}, we plot the `phase' diagram of the disorder parameter $\phi$ as a function of $\sigma$ and $\rho$. A line of discontinuous transitions is clearly observed at intermediate densities for all `temperatures' $\sigma$ \emph{above} a critical value, while the homogeneous state remains stable at both low (gas-like) and high (liquid-like) densities. Amusingly, in this liquid-gas analogy, the gas phase begins to `boil' as the temperature $\sigma$ increases—a counterintuitive phenomenon reminiscent of a paradox found in simple crowd models \cite{helbing2000freezing}.
Finally, we use the image of a liquid-gas transition to shed a different light on the hysteresis observed in Fig.~\ref{fig1}.
When $\sigma$ is varied at fixed density, as considered in Fig.~\ref{fig1}, the system moves along a vertical line in Fig.~\ref{fig:PhaseDiag}: it does not transit between two pure states, but from a pure phase into a co-existence region. Starting from the uniform regime, the system may avoid developing an instability until it hits the spinodal. In contrast, starting from the co-existence region, stop-and-go waves may persist up to a different neutral stability line, the binodal.
Interestingly, unlike in a liquid-gas mixture below the binodal line, there is no \emph{static} coexistence between the jammed and freely flowing phases in a steady state; instead, the jam moves continuously. Similar transitions into circulating states were recently identified in systems with non-reciprocal interactions \cite{fruchart2021non}; this is easily rationalized by considering that each car is tied by a spring to the predecessor, 
but not to its follower, 
so that the in and out currents cannot compensate at an interface.

\vspace{0.5cm}

In conclusion, we have shown that the SATG model can undergo an abrupt transition into a state characterized by stop-and-go waves similar to experiments \cite{tadaki2013phase}. Interestingly, the model is unconditionally stable and the transition is not the result of a linear 
instability but rather the consequence of the continuous action of the noise, which (1) destabilizes the homogeneous state, (2) stabilizes the heterogeneous state and (3) pushes the system between states. This is reminiscent of a liquid-gas phase transition (albeit with an unsteady phase co-existence), as qualitatively discussed for previously evidenced instabilities, notably in deterministic car-following models with a reaction time \cite{nagatani1998delay} (where the reaction time plays the role of the
temperature) or in the Nagel-Schreckenberg cellular automaton \cite{ke2004noise} and its continuous time-limit \cite{krauss1996continuous,jost2005probabilistic}, where at each time step cars may brake by a random amount (relative to the desired acceleration). Note, in passing, that this picture of a first-order transition need not be universal across all forms of traffic; for instance, pedestrian single-file traffic displays stop-and-go waves that are rather evanescent, considerably smaller than in highway traffic \cite{Tordeux2016b,EilhardtS15}, and not interspersed with freely flowing phases, unlike road traffic.
Coming back to the SATG instability, we have shown that a fruitful analogy can be drawn with Kapitza’s inverted pendulum, allowing for a quantitative stability analysis. 
Therefore, unlike other instabilities, the local properties of the response function $F$ around the stationary flow do not govern the instability. Instead, the variables that elude a deterministic description (in particular, the `human error' \cite{laval2014parsimonious}, i.e., the inaccurate drivers' perceptions or responses, their variability) play a central role.
Since macroscopic observations are insufficient to discriminate between the mechanisms, we are currently planning virtual reality experiments in which these physical ingredients can be separately varied to identify the key ingredients responsible for stop-and-go waves.

\newpage


\section*{End Matter}

\paragraph*{Simulation settings.}

We simulate $N$ cars on a single lane of length $L$ with periodic boundary conditions. By default, we set $N=22$ and $L=231\,\mathrm{m}$ to mimic the experiments of \cite{sugiyama2008traffic,tadaki2013phase} at the onset of traffic oscillations. Cars are ordered from $n=1$ to $n=N$ ($n=1$ is the predecessor of $n=N$ because of the periodic boundaries); overtaking is precluded. At time $t$, car $n$ is at position $x_n(t)$ 
and moves at speed $v_n(t)=\dot x _n(t)$.
The inter-vehicular spatial gap $\Delta x_n(t)$ and the speed difference $\Delta v_n(t)$ are given by 
$$
\begin{cases}
~\Delta x_n(t)=x_{n+1}(t)-x_n(t)-\ell,\qquad n\in\{1,\ldots,N-1\},\\[1mm]
~\Delta x_N(t)=L+x_1(t)-x_N(t)-\ell,
\end{cases}
$$
where $\ell=5$~m is the car length, and
$$
\begin{cases}
~\Delta v_n(t)=v_{n+1}(t)-v_n(t),\qquad n\in\{1,\ldots,N-1\},\\[1mm]
~\Delta v_N(t)=v_{1}(t)-v_N(t),
\end{cases}
$$
respectively. The time dependencies will be dropped in the following.

\vspace{0.5cm}
\paragraph*{Car-following models.}
The following models have been implemented:
\begin{itemize}
    \item the Stochastic Full Velocity Difference (SFVD), as a representative of the class of (linear or near-linear) optimal velocity and full velocity difference models. It is given by \cite{wagner2011time,treiber2009hamilton,wang2020stability,friesen2021spontaneous}
\begin{eqnarray}
    \mathrm{d}v_n=\bigg(\frac{V\bigl(\Delta x_n\bigr)-v_n}{T_1}+\frac{\Delta v_n}{T_2}\bigg)\mathrm{d}t+\sigma \mathrm{d} W_n
    \label{SFVDMapp}
\end{eqnarray}
where $T_1=2.5$~s and $T_2=2$~s are two relaxation times and where $V:\mathbb R\mapsto\mathbb R_+$ is the optimal velocity (OV) function given by the sigmoid 
\begin{eqnarray}
    V(s)=v_0\frac{\tanh(s/\ell_0-\kappa)+\tanh(\kappa)}{1+\tanh(\kappa)},
\end{eqnarray}
where $\kappa=0.5$ and $\ell_0=20$~m are the shape and scale parameters, and where $v_0=20$~m/s is the desired speed \cite{bando_DynamicalModelTraffic_1995,jiang_FullVelocityDifference_2001,treiber2009hamilton}.

\item the stochastic (near-linear) model of Tomer et al.  \cite{tomer2000presence} given by
\begin{eqnarray}
    \mathrm{d}v_n &= K \bigg(1 - \frac{2v_n T + \ell}{\Delta x_n+\ell}\bigg)\mathrm{d}t + \frac{Z^2\bigl(-\Delta v_n\bigr)}{ 2\Delta x_n}\mathrm{d}t \nonumber 
    \\ &-2Z\bigl(v_n - v_0\bigr)\,\mathrm{d}t + \sigma \mathrm{d} W_n
    \label{Tomerapp}
\end{eqnarray}
where $Z(x)=(x+|x|)/2$ is the positive part of $x$, $K=5$~m/s$^2$ is a sensitivity parameter, and $T=1$~s is the desired time gap. 

\item the Stochastic Intelligent Driver Model (SIDM) \cite{treiber2017intelligent}, which ) reads 
\begin{eqnarray}
\begin{array}{l}
    \displaystyle \mathrm{d}v_n = a \bigg( 1 - \bigg(\frac{f\bigl(v_n,\Delta v_n\bigr)}{\Delta x_n}\bigg) ^ 2 - \bigg(\frac{v_n}{v_0}\bigg) ^ 4\bigg)\mathrm{d}t + \sigma \mathrm{d}W_n\\[4mm]
\displaystyle \text{with}\quad  f(v,\Delta v) = s_0 + T v - v \frac{\Delta v}{2\sqrt{ab}},
\end{array}
\label{SIDMapp}
\end{eqnarray}
where $a=b=2$~m/s$^2$ are the desired acceleration and maximal deceleration parameter,  $s_0=2$~m is a minimal gap, $T=1$~s is the desired time gap and where $v_0=20$~m/s is the desired speed.

\item the Stochastic Adaptive Time Gap (SATG) model, obtained by relaxing the time gap $T_n(t)=\Delta x_n/v_n$ as $\dot T_n(t)=\lambda(T-T_n(t))$, where $\lambda=0.2$~s$^{-1}$ is a sensitivity parameter and $T=1$~s is the desired time gap \cite{tordeux2010adaptive}. Using speed and gap variables, the model reads
\begin{eqnarray}
    \mathrm{d}v_n =\frac{\lambda \left(\Delta x_n-Tv_n\right)+\Delta v_n}{T_\varepsilon\big(\Delta x_n,v_n\big)}\mathrm{d}t+\sigma\mathrm{d}W_n,
    \label{SATGapp}
\end{eqnarray}
For practical purposes (namely, to avoid singularities when the cars are about to collide due to the noise or when their speed goes to zero), the time gap in the denominator is bounded  between $\tmin=0.1$~s and $\tmax=4$~s using the mollifier 
\begin{eqnarray}
        T_\varepsilon(\Delta,v)= f_\varepsilon\left(\tmin,f_{-\varepsilon}\left(\tmax,\frac{\Delta}{f_{\varepsilon}(0,v)}\right)\right),
    \label{SATGapp2}
\end{eqnarray}
where $f_\varepsilon$ is the {LogSumExp} function $
   f_\varepsilon(a,b)=\varepsilon \log\big(e^{a/\varepsilon}+e^{b/\varepsilon})$.
The function $f_\varepsilon(a,b)$ converges to the maximum of $a$ and $b$ as $\varepsilon\to0^+$ and to the minimum as $\varepsilon\to0^-$. In practice, $\varepsilon$ is set to 0.01.

\end{itemize}

In all the above expressions,  $\sigma \mathrm{d}W_n$ represents the stochastic noise, where the ${W_n(t)}$ denote independent Wiener processes (see below for the actual implementation). The deterministic versions of the models are obtained by setting the volatility $\sigma$ to 0.


\vspace{0.5cm}
\paragraph*{Numerical implementation.}
We used an implicit/explicit Euler--Maruyama numerical solver for the simulation with a time step $\delta t=0.001$~s.
for the $n$-th agent, $n\in\{1,\ldots,N\}$, the numerical scheme reads
\begin{eqnarray}\label{eq:numsch}
    \begin{cases}
    ~\displaystyle x_n(t+\delta t)=x_n(t)+\delta t\, v_n(t+\delta t)\\[1mm]
    ~\displaystyle v_n(t+\delta t)=v_n(t)+\delta t\, F_n(t) +\sqrt{\delta t}\,g(v_n(t)) \xi_n(t)
    \end{cases}
\end{eqnarray}
where $F_n(t)$ is used as a shorthand for the model-dependent deterministic response function $F\bigl(\Delta x_n(t),v_n(t),v_{n+1}(t)\bigr)$, the $\xi_i(t)$'s (for $i=1,\ldots,N$ and $t\in\delta t\mathbb N$) are independent normal random variables, and
\begin{eqnarray}
    g(v)=\frac{\sigma}{1+\exp(-\alpha(v-v_\sigma))},\quad \sigma\ge0,~\alpha=10^3.~~~~~
\end{eqnarray}
 is designed to be close to zero when $v$ gets smaller than $v_\sigma=0.1$~m/s to limit the collisions and equal to the volatility constant $\sigma$ when  $v>\!\!>v_\sigma$.

\vspace{0.5cm}
\paragraph*{Distribution of times for the emergence of stop-and-go waves in SATG.}

\begin{figure*}[!ht]
    \centering
    \includegraphics[width=\textwidth]{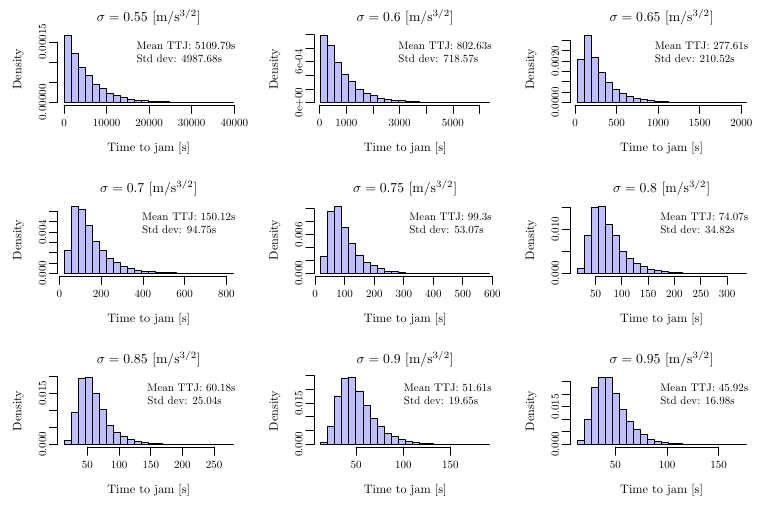}\vspace{-2.5mm}
    \caption{Histograms of the time for a stop-and-go wave to appear (TTJ) for a system with random initial conditions for noise volatilities ranging from 0.55 to 0.95~m/s$^{3/2}$. The more volatile the noise, the faster the waves appear.}
    \label{fig:TTJ}
\end{figure*}

In the ring experiment, the time for a stop-and-go wave to emerge is about 150~s in \cite{sugiyama2008traffic} and 320~s in \cite{nakayama2009metastability}. 
These times, called time-to-jam (TTJ), are also variable in the SATG model~\eqref{SATGapp}-\eqref{SATGapp2}. 
Figure~\ref{fig:TTJ} shows the histograms of the TTJ for noise volatilities $\sigma$ ranging from 0.55 to 0.95~m/s$^{3/2}$. 
$10^4$ simulations are repeated for each $\sigma$.
The initial condition is uniform. 
We consider that a stop-and-go wave has emerged when the standard deviation of gaps exceeds 6~m. 
The more volatile the noise, the faster the waves appear. 
The empirically observed variabilit   seems to correspond to $\sigma\ (\mathrm{m/s^{3/2}}) \in [0.6,0.7]$.

\vspace{0.5cm}
\paragraph*{Computation of $\langle \phi \rangle_t$.}
In Fig.~\ref{fig1}, we run warm-up simulations for 5,000 seconds (to reach stationarity) before time-averaging the gap standard deviation $\phi$
over the next 2000 seconds. We repeat 100 independent Monte Carlo simulations for each value of the noise volatility ranging from 0 to 1 in steps of 0.02 and each of the four stochastic car-following models of Eqs.~\eqref{SFVDMapp}--\eqref{SATGapp}. 
In the figure, the curves are the mean values of the Monte Carlo simulations, while the colored areas show the minimum/maximum range of variation.

\vspace{0.5cm}
\paragraph*{Online simulation platform}
An online simulation platform of the solver \eqref{eq:numsch} for the setting of the experiment by Sugiyama et al.\ \cite{sugiyama2008traffic} and the stochastic car-following models of  Eqs.~\eqref{SFVDMapp}--\eqref{SATGapp} is available at
\urllink{https://www.vzu.uni-wuppertal.de/fileadmin/site/vzu/Experiment_by_Sugiyama_et_al._2007.html?speed=0.8}{https://www.vzu.uni-wuppertal.de/fileadmin/site/vzu/Experiment\_by\_Sugiya- ma\_et\_al.\_2007.html?speed=0.8}

\clearpage

\begin{center}
    {\LARGE \bf Supplementary Material} 
\end{center}
\def\ge{g_\text e} 
\def\gen{g_{{\rm e},n}}
\def\ve{v_\text e}
\def\d{\text d}
\def\bb{\langle b\rangle}
\def\bia{\langle 1/a\rangle}

\def\fgn{f_n^g}
\def\fvn{f_n^v}
\def\fdvn{f_n^{\Delta v}}
\def\fg{f_g}
\def\fv{f_v}
\def\fdv{f_{\Delta v}}

\def\tmin{T_{\rm min}}
\def\tmax{T_{\rm max}}


 \setcounter{equation}{0}
 \def\theequation{S\arabic{equation}}
 \setcounter{figure}{0}
 \def\thefigure{S\arabic{figure}}

\section*{A. ~Numerical simulations of the SATG model} \label{AppA}

\subparagraph*{Fine-tuning SATG model to reproduce the stop-and-go waves observed in Sugiyama's experiment}

\begin{figure}[!ht]
    \includegraphics[width=.75\textwidth]{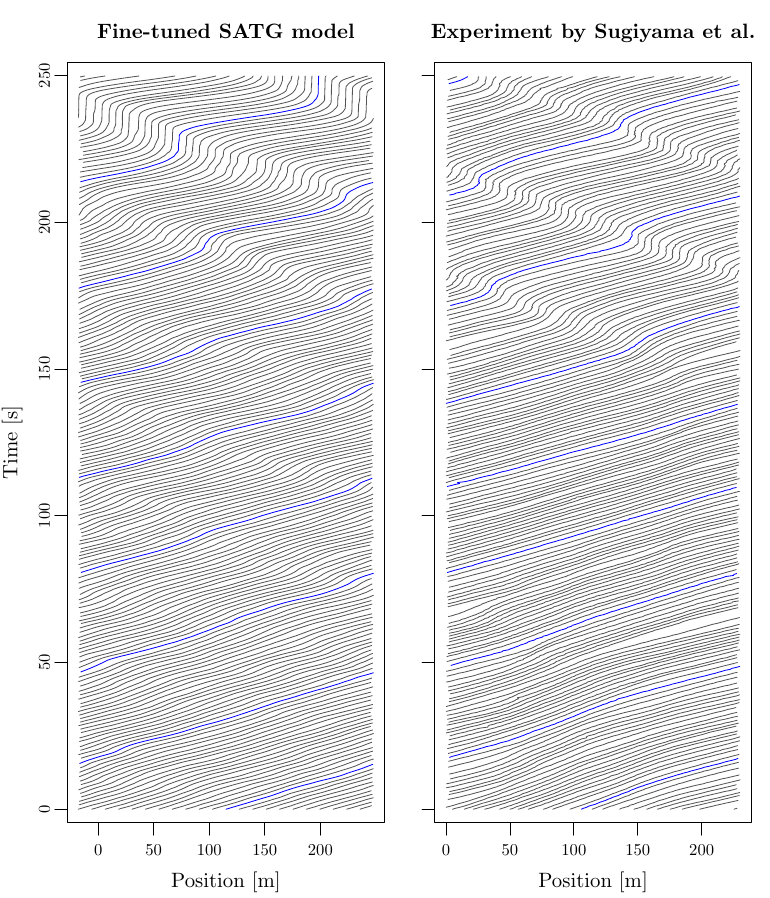}
    \caption{Trajectories, featuring stop-and-go waves with a selected simulation of the fine-tuned SATG model with $T=0.7$~s, $\lambda=1/3$~s$^{-1}$, $\tmax=2$~s and $\sigma=1.2$~m/s$^{3/2}$ (left panel) and in the experiment of Sugiyama et al. \protect\cite{sugiyama2008traffic} (right panel).}
    \label{fig:traj2}
\end{figure}

Figure~\ref{fig:traj2} shows the simulated trajectories with a fine-tuned SATG model (Eq.~9-10) with $\sigma=1.2$~m/s$^{3/2}$ that better reproduces the stop-and-go dynamics observed in Sugiyama's experiment \cite{sugiyama2008traffic}. 
The time gap is set to $T=0.7$~s in order to match the task given to the drivers during the experiment (driving at a speed of 30 km/h, which corresponds to a time gap of 0.7 seconds, given the density). In addition, compared to the values used in the main text, the relaxation rate $\lambda=1/3$~s$^{-1}$ was increased and the maximum time gap $\tmax=2$~s was decreased to reduce the duration of the stopping phase.

\subparagraph*{Bimodality at the transition}

\begin{figure}[!ht]
    \centering\medskip
    \includegraphics[width=.45\textwidth]{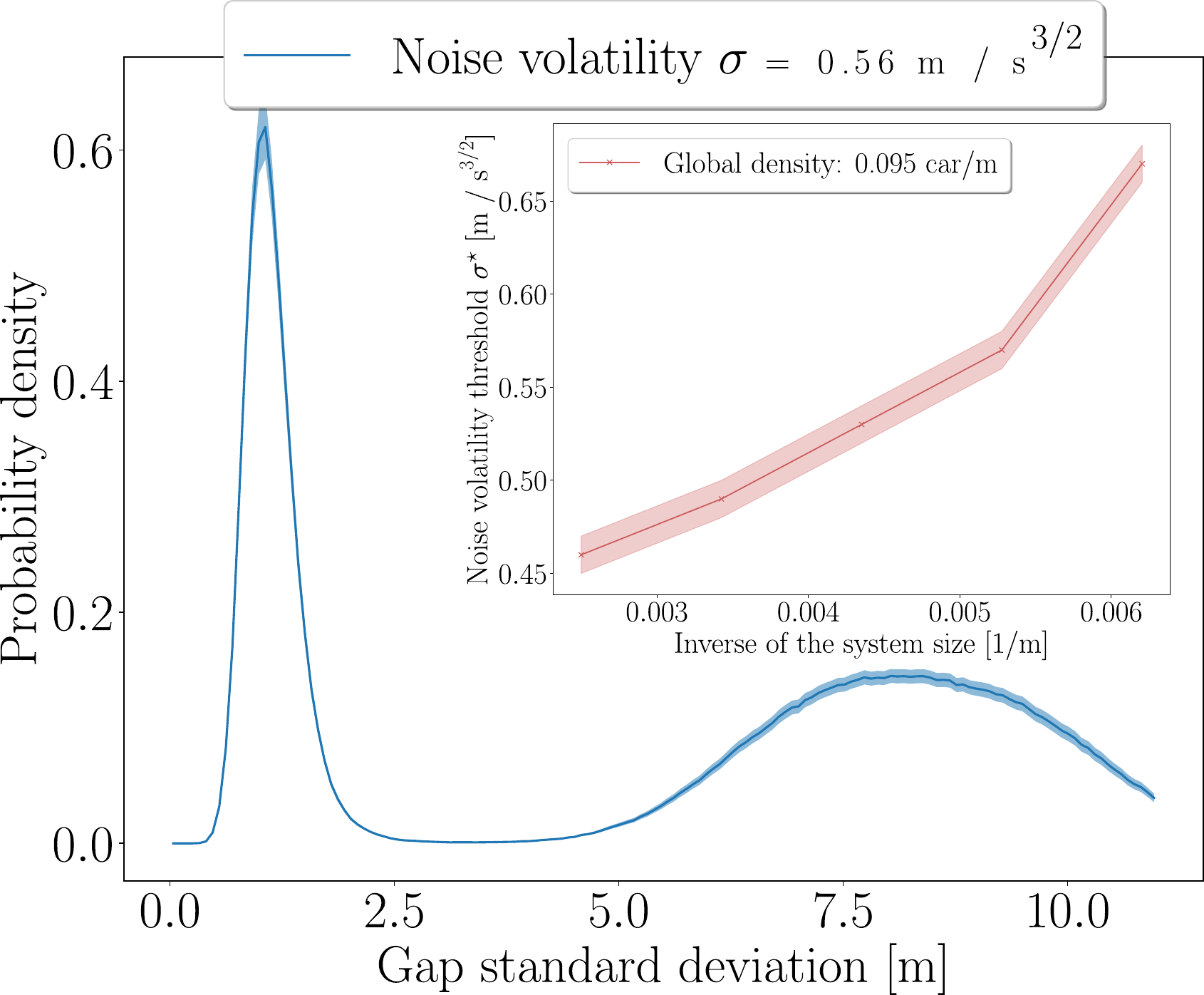}
    \caption{Probability density function of the gap standard deviation $\phi$ over time and over independent simulations 
    of the single-file periodic system with $N=22$ cars close to the transition $\sigma=\sigma^{\star}$. \emph{Inset:} Variation of the transition threshold $\sigma^{\star}$ with the (inverse) system size.}
    \label{fig:bimodality}
\end{figure}

Figure~\ref{fig:bimodality} presents the distribution of the standard deviation of the gap at the critical volatility $\sigma=0.56$~m/s$^{3/2}$ where the system is metastable (see Fig.~4). 
The system oscillates between a homogeneous state and stop-and-go dynamics at this noise level and the gap standard deviation  has a bimodal distribution.

\section*{B. ~Linear stability analysis of the biased deterministic model}\label{AppB}

\def\ge{g_\text e}
\def\gen{g_n^\text e}
This Appendix is dedicated to the linear stability analysis of the SATG model, either around the uniform stationary flow or around a heterogeneous state.

To explore the system's dynamics in the presence of heterogeneities, we bias the accelerations of the SATG model (Eq.~9-10) with time-independent offset terms $b_n$ ($n\in\{1,\ldots,N\}$), as follows:
\begin{equation}
    \dot v_n(t)=F\bigr(v_n(t),\Delta x_n(t),\Delta v_n(t)\bigr)+b_n,
    \label{eq:supp_deter_bias}
\end{equation}
where $v_n(t)$ is the speed, $\Delta x_n(t)$ is the gap, $\Delta v_n(t)$ is the speed difference to the predecessor and where
\begin{equation}\label{eq:ATGbias}
    F(v,\Delta x,\Delta v)=\lambda v \left(1-\frac{Tv}{\Delta x}\right) + \frac{v\Delta v}{\Delta x},
\end{equation}
with $\lambda>0$ the sensitivity parameter, $T$ the desired time gap parameter, and $b_n$ a constant and vehicle specific bias in the acceleration.

\subparagraph*{Equilibrium solution}
We consider $N\geq 2$ vehicle of length $\ell\geq0$ on a ring of length $L>N\ell$. 
The equilibrium gap for the homogeneous system where $b_n=0$ for all $b\in\{1,\ldots,N\}$ is given by 
\begin{equation}
    \ge=L/N-\ell.
\end{equation}
In the presence of acceleration biases, the equilibrium solution for which $\dot v_n=0$ for all $n\in\{1,\ldots,N\}$ is not uniform in space. 
This is the $\big(\ve,(\gen)_{n=1}^N\big)$ configuration satisfying
\begin{equation}
\left\{\begin{array}{l}
     \displaystyle\sum_{n=1}^N \gen=L-N\ell=N\ge,\\[5mm]
     F(\gen,\ve,0)+b_n=0,\qquad n\in\{1,\ldots,N\}.
\end{array}\right.
\end{equation}
We can deduce from the second part that
\begin{equation}\label{eq:gen}
    \gen=\frac{\lambda T\ve^2}{b_n+\lambda \ve},\qquad\forall n\in\{1,\ldots,N\},
\end{equation}
while, using the conservation of spacing $\sum_n \gen=N\ge$, the equilibrium speed becomes the solution of
\begin{equation}\label{eq:ve}
    \sum_{n=1}^N \frac{\lambda T\ve^2}{b_n+\lambda \ve}=L-N\ell.
\end{equation}
Note that the equilibrium gaps \eqref{eq:gen} are positive for all the vehicles if
\begin{equation}\label{eq:c1ATG}
    \ve> -\frac1\lambda\min_n b_n.
\end{equation}
In addition, we recover the equilibrium solution of the homogeneous ATG model
\begin{equation}
\ve=\frac{\ge}{T}\qquad\text{and}\qquad
\gen=\ge~\text{ for all }~n\in\{1,\ldots,N\},
\end{equation} 
if the biases are zero, i.e., $b_n=0$ for all $n\in\{1,\ldots,N\}$.
Furthermore, in the case where the bias $b_n=b$ is identical for all vehicles $n\in\{1,\ldots,N\}$, we have 
\begin{equation}
    \lambda T\ve^2-(b+\lambda \ve)\ge=0,
\end{equation}
and we can deduce that
\begin{equation}\label{eq:veATG2}
    \ve=\frac{\ge\lambda+\sqrt{(\ge\lambda)^2+4\lambda Tb\ge}}{2\lambda T}
    =\frac{\ge}{2T}\left(1+\sqrt{1+\frac{4Tb}{\lambda\ge}}\right)
\end{equation}
The equilibrium speed exists if $1+\frac{4Tb}{\lambda\ge}\geq0$ and we obtain the condition
\begin{equation}\label{eq:c2ATG}
    b\geq -\frac{\lambda\ge}{4T}.
\end{equation}
Note that \eqref{eq:c2ATG} implies the preliminary condition $b>-\lambda\ve$ (see \eqref{eq:c1ATG}).

\subparagraph*{Linearization of the system}
\def\fgn{f_n^{\Delta x}}
\def\fvn{f_n^v}
\def\fdvn{f_n^{\Delta v}}
The partial derivatives of the model \eqref{eq:ATGbias} at equilibrium are given by
\begin{equation}
    \fgn=\frac{\partial F}{\partial g}(\gen,\ve,0)= 
    \frac{\lambda T\ve^2}{(\gen)^2},
    \qquad
    \fvn=\frac{\partial F}{\partial v}(\gen,\ve,0)= 
    \lambda\left(1-\frac{2T\ve}{\gen}\right),
    \quad\text{and}\quad
    \fdvn=\frac{\partial F}{\partial \Delta v}(\gen,\ve,0)= 
    \frac{\ve}{\gen}\\
\end{equation}
The characteristic equation of the linearised system is
\begin{equation}\label{eq:EC}
    \prod_{n=1}^N \left[z^2-z(\fvn-\fdvn)+\fgn\right]-\prod_{n=1}^N \left[z\fdvn+\fgn\right]=0,\qquad\theta\in[0,2\pi].
\end{equation}
A general sufficient linear stability condition of a heterogeneous model for which all roots of \eqref{eq:EC} have strictly negative real parts except one equal to zero (due to periodic boundary conditions) is \cite[Eq.~(5)]{ngoduy2015effect}
\begin{equation}\label{eq:SC}
    \sum_{n=1}^N \left[\frac12\left(\frac{\fvn}{\fgn}\right)^2-\frac{\fvn\fdvn}{\fgn\fgn}-\frac1{\fgn}\right]\geq0.
\end{equation}
We have
\begin{equation}
    \frac{\fvn}{\fgn}
    =\frac{\lambda\left(1-\frac{2T\ve}{\gen}\right)}{\frac{\lambda T\ve^2}{(\gen)^2}}
    =\frac{\gen(\gen-2T\ve)}{T\ve^2},
\end{equation}
while
\begin{equation}
    \frac{\fvn\fdvn}{\fgn\fgn}
    =\frac{\lambda\left(1-\frac{2T\ve}{\gen}\right)\frac{\ve}{\gen}}{\frac{\lambda^2 T^2\ve^4}{(\gen)^4}}
    =\frac{(\gen)^2(\gen-2T\ve)}{\lambda T^2\ve^3}.
\end{equation}
The sufficient linear stability condition \eqref{eq:SC} is then given by
\begin{equation}
    \sum_{n=1}^N \left[\frac12\left(\frac{\gen(\gen-2T\ve)}{T\ve^2}\right)^2
    -\frac{(\gen)^2(\gen-2T\ve)}{\lambda T^2\ve^3}
    -\frac{(\gen)^2}{\lambda T\ve^2}\right]\geq0,
\end{equation}
or again
\begin{equation}
    \sum_{n=1}^N\;(\gen)^2 \left[\frac{\lambda(\gen-2T\ve)^2}{2T\ve^2}
    -\frac{\gen-2T\ve}{T\ve}-1\right]
    =\sum_{n=1}^N\;(\gen)^2 \left[\frac{\lambda(\gen-2T\ve)^2}{2T\ve^2}
    -\frac{\gen-T\ve}{T\ve}\right]\geq0.
\end{equation}
Then, using
$\gen=\frac{\lambda T\ve^2}{b_n+\lambda \ve}$ 
and remarking that
$\gen-2T\ve=-T\ve\frac{2b_n+\lambda \ve}{b_n+\lambda \ve}$
while
$\gen-T\ve=-T\ve\frac{b_n}{b_n+\lambda \ve}$,
we obtain 
\begin{equation}
    \sum_{n=1}^N\;\left(\frac{\lambda T\ve^2}{b_n+\lambda \ve}\right)^2 
    \left[\frac{\left(T\ve\frac{2b_n+\lambda \ve}{b_n+\lambda \ve}\right)^2}{2T\ve^2}
    +\frac{T\ve\frac{b_n}{b_n+\lambda \ve}}{T\ve}\right]
    \geq0.
\end{equation}
After simplifications (we have $\lambda,T,\ve>0$), it follows
\begin{equation}
    \sum_{n=1}^N\;\frac{1}{(b_n+\lambda \ve)^2}
    \left[\frac{\lambda T}2\left(\frac{2b_n+\lambda \ve}{b_n+\lambda \ve}\right)^2
    +\frac{b_n}{b_n+\lambda \ve}\right]
    \geq0,
\end{equation}
or again
\begin{equation}\sum_{n=1}^N\;\frac{1}{(b_n+\lambda \ve)^4}
    \left[\frac{\lambda T}2\left(2b_n+\lambda \ve\right)^2
    +b_n(b_n+\lambda \ve)\right]
    \geq0.
\end{equation}
We have
\begin{equation}\label{eq:intermed}
\begin{array}{lcl}
    \displaystyle\frac{\lambda T}2\left(2b_n+\lambda \ve\right)^2 + b_n(b_n+\lambda \ve)
    &=&\displaystyle2\lambda Tb_n^2+\frac12\lambda^3T\ve^2+2\lambda^2Tb_n\ve+b_n^2+b_n\lambda\ve\\[2mm]
    &=&\displaystyle b_n^2(2\lambda T+1)+b_n(2\lambda T+1)\lambda\ve+\frac12\lambda^3T\ve^2\\[2mm]
    &=&\displaystyle b_n(2\lambda T+1)(b_n+\lambda\ve)+\frac12\lambda^3T\ve^2,
\end{array}
\end{equation}
and the linear stability condition can be written
\begin{equation}\sum_{n=1}^N\;\frac{b_n(2\lambda T+1)}{(b_n+\lambda \ve)^3}
+\frac{\lambda^3T\ve^2}{2(b_n+\lambda \ve)^4}
    \geq0.
\end{equation}

Since $\lambda,T>0$, the model is linearly stable if $b_n=0$ for all $b_n\in\{1,\ldots,N\}$. 
Indeed the homogeneous ATG model is unconditionally linearly stable \cite{khound2021extending}.
In addition, when all the biases are identical, i.e., $b_n=b>-\lambda\ge/(4T)$ for all $b_n\in\{1,\ldots,N\}$, dividing by $\lambda T\ve^2>0$ the last line of \eqref{eq:intermed} and using \eqref{eq:veATG2}, we obtain the linear stability condition of the biased ATG model 
\begin{equation}
    \frac{b(2\lambda T+1)}{\ge}+\frac{\lambda^2}{2}\geq0.
\end{equation}
This is
\begin{equation}
    b\geq\frac{-\lambda^2\ge}{4\lambda T+2}.
\end{equation}
The bias has to be negative and sufficiently low, especially for high $\lambda$ or $\ge$ (i.e., low density), to destabilise the system.

\section*{C. ~Extended stability analysis of the periodically driven system}
\label{AppC}
This Section details the extended stability analysis of the car-following system in which the noise term is substituted by an externally applied deterministic oscillatory driving, with vanishing residual noise:
\begin{equation}
 \dot{v}_n(t)=F\big(\Delta x_n(t),v_n(t),v_{n+1}(t)\big) + C\,\cos(\omega t + \varphi_n)\label{eq:cyclic_basic2}.
\end{equation}

We assume that the driving frequency is low enough so that a pseudo-stationary approximation can be performed, i.e., at each time $t$ the system follows the biased deterministic equation of Eq.~\eqref{eq:supp_deter_bias} with $b_n = \,\cos(\omega t + \varphi_n)$. Perturbations around the  base flow grow at rates $\nu$ given by the real parts of the eigenvalues $z$ in Eq.~\eqref{eq:EC}. Unfortunately, finding the roots $z$ of this equation is far from straightforward analytically. The equation is thus solved numerically, first by deducing the equilibrium speed from Eq.~\eqref{eq:ve} and the associated gaps, and then finding the nontrivial complex roots $z\neq 0$ (the eigenvalue $0$ is not relevant physically) of  Eq.~\eqref{eq:EC} using a Newton-Raphson method  with multiple starting points located on a regular lattice in complex space; we check that no eigenvalue has been missed by quadrupling the number of starting points. Numerically, identifying that $z$ is a root of Eq.~\eqref{eq:EC} can be challenging for small $\lambda$ or large $T$ because a quasi-continuum of $z$ values yield vanishingly small, but nonzero products. 
Accordingly, the validity of candidate roots is checked by calculating the ratio of the two products in Eq.~\eqref{eq:EC}. We denote by $\nu\bigl(\{b_n\}\bigr)= \max\ \Re e(z)$ the largest growth rate over all roots $z$.

\begin{figure}[!ht]
    \centering\medskip
    \includegraphics[width=.45\textwidth]{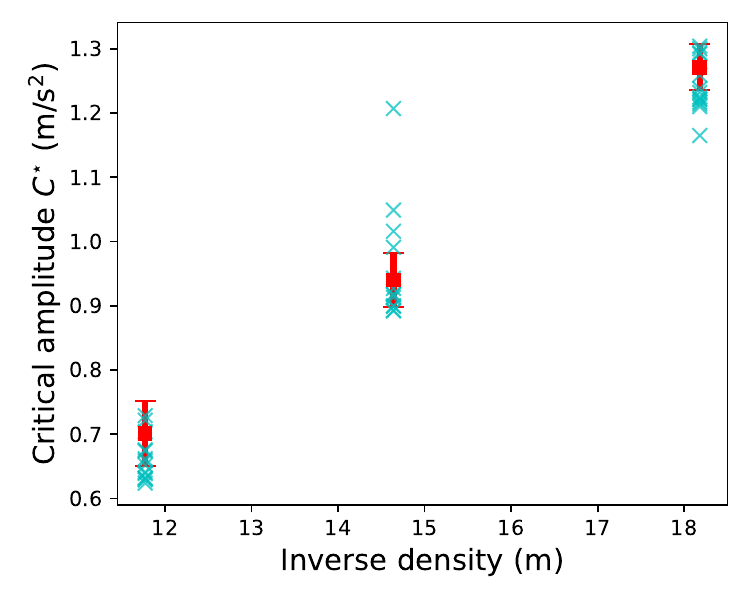}
    \caption{Critical amplitudes $C^{\star}$ predicted by our extended stability analysis, using the time-dependent phases of Eq.~\eqref{eq:ext_stab_cos} (cyan crosses) or a random distribution of phases $\theta_n$ (red squares)
    at different inverse densities, for $\lambda=0.2$ and $T=1$ and $N=22$ cars, with various ring perimeters. The error bars in red represent the standard errors, while the different crosses at a given density correspond to different initial phases ${\varphi_n}$.}
    \label{fig:random_phase_approx}
\end{figure}

If the eigenmode associated with this growth rate is roughly the same throughout the cycle (which is reasonable, because it tends to be the mode with the largest possible wavelength in the system), then the fastest growing perturbation (under our pseudo-stationary assumption) will unfurl as
\begin{equation}
 \exp\biggl(\int_0^{2\pi} \frac{d \theta}{\omega} \nu\Big( \{ C\ \cos(\theta + \varphi_n)\}\Big)\biggr)
 \label{eq:ext_stab_cos}
\end{equation}
over a cycle, hence an effective growth rate 
 \begin{equation}
     \nu_{\mathrm{eff}} (C) = \langle \nu(\{C\, \cos(\theta_n)\}) \rangle_{\theta},
 \label{eq:nu_eff}
 \end{equation}
 where $\theta_n$ is used as a shorthand for $\theta + \varphi_n$ and the angular brackets denote an average over $\theta\in [0,2\pi[$. We surmise that averaging $\nu_{\mathrm{eff}} (C)$ over random, uniformly distributed phases $\varphi_n$ is tantamount to averaging it over uniformly distributed $\theta_n$ in $\mathbb{R}/2\pi\mathbb{Z}$. Numerical simulations confirm that this is a decent approximation for an arbitrary set of phases $\{\varphi_n\}$ and quite a satisfactory one upon averaging over the $\{\varphi_n\}$, as shown in Fig.~\ref{fig:random_phase_approx}.

\section*{D. ~Correspondence between the critical noise volatility $\sigma^{\star}$ and the critical driving amplitude $C^{\star}$}

In the main text, we showed that the instability that emerges in SATG above a critical noise level $\sigma^{\star}$ (volatility)  is mirrored by an instability in the deterministic ATG model subject to oscillatory driving $C\,\cos(\omega t + \varphi_n)$, when $C\geqslant C^{\star}$ ($C^{\star}$ is virtually insensitive to the frequency $\omega$, provided it is low enough, typically below $0.1\,\mathrm{rad/s})$; the latter instability is rationalized in Appendix~{\bf C}. This Appendix delves into the relation between the critical amplitudes $\sigma^{\star}$ and $C^{\star}$.

\subparagraph*{A density-independent coefficient relating $\sigma^{\star}$ and $C^{\star}$.}
First, we provide some numerical evidence for the existence of a density-independent coefficient $A(\lambda,T)>0$ such that $C^{\star}=A(\lambda,T)\,\sigma^{\star}$, where $A(\lambda,T)$ may depend on model parameters $\lambda$ and $T$, but not on the car density. Note that $\sigma^{\star}$ and $C^{\star}$  do not have the same units; $A(\lambda,T)$ has dimension $\mathrm{s^{-1/2}}$.

\begin{figure}[!ht]
    \centering\medskip
    \includegraphics[width=.45\textwidth]{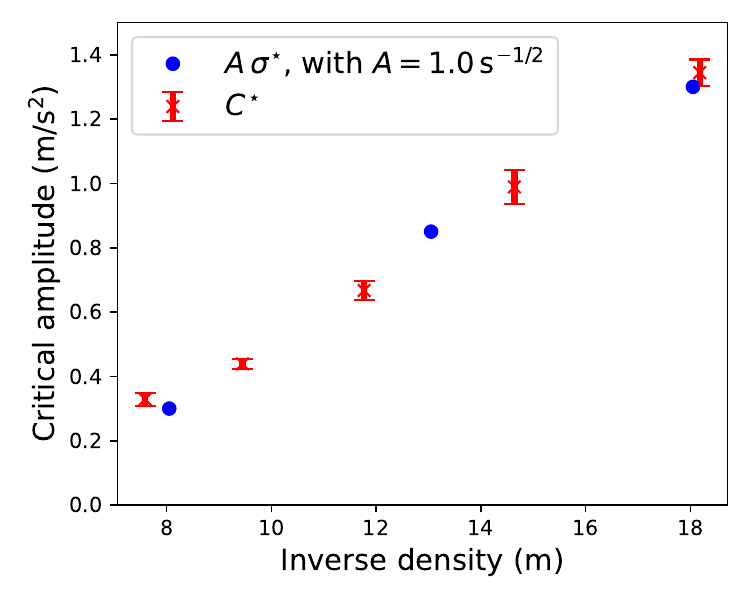}
    \caption{Critical amplitudes $\sigma^{\star}$ and $C^{\star}$ at different inverse densities, for the SATG model parameters $\lambda=0.2$ and $T=1$ used in the main text. The critical volatility $\sigma^{\star}$ is obtained by direct numerical simulations of SATG, whereas $C^{\star}$ is calculated on the basis of the extended stability analysis presented in Appendix~{\bf C}. The coefficient $A(\lambda,T)=1.0\,\mathrm{s^{-1/2}}$ was adjusted manually.}
    \label{fig:sigma_vs_C}
\end{figure}

Figure~\ref{fig:sigma_vs_C} supports the existence of $A(\lambda,T)$ and its independence of density for the model parameters used in the main text. The same conclusion is reached for distinct model parameters (provided that numerical simulations of SATG and the extended stability analysis both point to a clear critical amplitude, which is not always the case when one significantly departs from the baseline values).

\subparagraph*{Analytical derivation of the coefficient $A(\lambda,T)>0$ around the steady state.}

Under asynchronous (i.e., $\varphi_n\neq \mathrm{cst}$) oscillatory driving \footnote{Note that an instability is also found for synchronous oscillations $\varphi_n= \mathrm{cst}$, but at \emph{much} higher $\sigma$.}, for $C\geqslant C^{\star}$, the instability arises because the system is pushed to a (possibly, quasisteady) heterogeneous state which is linearly unstable (see Appendix~{\bf C}). The heterogeneous gaps $\Delta x_n$  thus have a central role in triggering the instability. Accordingly, we strive to estimate the coefficient $A(\lambda,T)$ on a theoretical basis by equating the mean-square gap fluctuations $ \langle |\Delta x|^2 \rangle$ induced by white noise, on the one hand, and by oscillatory driving, on the other hand, around the steady-state uniform flow. Below, we derive detailed expressions for these gap fluctuations, but simpler approximate formulae can be obtained by reasoning on a harmonic oscillator subjected to either thermal fluctuations or periodic driving.

We start by linearizing the equation of motion (Eq.~\ref{eq:supp_deter_bias}), schematically written as
$\dot{v}_{n}=F_{n}(\Delta x_{n},v_{n},v_{n+1})+\Xi_{n}(t)$,
around the uniform flow, in the presence of a noise term $\Xi_n (t)$ (which will be either white noise or a sinusoidal driving). We denote the deviations  from the uniform flow positions and speeds  by $\delta x_n(t)$ and $\delta v_n (t)$, respectively. Considering one realization of the dynamics over the time window $t\in [0,T_{sim}]$, we conduct the linear expansion in Fourier space (with $\widehat{\delta x_n}(\omega)=\int_{-\infty}^{\infty} e^{-i\omega t} \delta x_n(t)dt $ and $\widehat{\delta v_n}=i\omega \widehat{\delta x_n}$),  as follows
\begin{equation}
-\omega^{2}\widehat{\delta x_{n}}=F_{n,\Delta x_{n}}\cdot\widehat{\delta\Delta x_{n}}+F_{n,v_{n}}\cdot i\omega\widehat{\delta x_{n}}+F_{n,v_{n+1}}\cdot i\omega\left(\widehat{\delta x_{n}}+\widehat{\delta\Delta x_{n}}\right)+\widehat{\Xi_{n}},
\end{equation}
where the partial derivatives read 
\begin{equation}
F_{n,\Delta x_{n}}=\frac{\lambda}{T};\,F_{n,v_{n}}=-\lambda-\frac{1}{T};\,F_{n,v_{n+1}}=\frac{1}{T}.
\end{equation}

This directly leads to
\begin{equation}
\mathcal{N}\,\widehat{\delta x_{n}}=\mathcal{P}\,\widehat{\delta\Delta x_{n}}+\widehat{\Xi_{n}},
\label{eq:linear}
\end{equation}
where $\mathcal{N}(\omega)=-\omega^{2}+i\omega\lambda$ and $\mathcal{P}(\omega)=T^{-1}\left(\lambda+i\omega\right)$ (we dropped the $(\omega)$ dependencies out of convenience). Subtracting the foregoing equation for $n$ from that for $n+1$ and grouping terms, we arrive at the following equation
\begin{equation}
\left(\mathcal{N}+\mathcal{P}\right)\widehat{\delta\Delta x_{n}}=\mathcal{P}\widehat{\delta\Delta x_{n+1}}+\left(\widehat{\Xi_{n+1}}-\widehat{\Xi_{n}}\right),
\end{equation}
whose square norm gives the mean-square gap fluctuations

\begin{equation}
\label{eq:eq5}
|\mathcal{N}+\mathcal{P}|^{2}\Vert \widehat{\delta\Delta x_{n}}\Vert ^{2}=\left|\mathcal{P}\right|^{2}\Vert \widehat{\delta\Delta x_{n+1}}\Vert ^{2}+\mathrm{CrCor}+(\left\Vert \widehat{\Xi_{n+1}}\Vert ^{2}+\left\Vert \widehat{\Xi_{n}}\right\Vert ^{2}\right),
\end{equation}
where we have used the shorthand $\left\Vert \bullet \right\Vert^2 = \langle |\bullet|^2\rangle$ for the average of the square norm over realizations. The second term on the right-hand side of Eq.~\eqref{eq:eq5}, $\mathrm{CrCor}=2\mathcal{R}e\left[\mathcal{P}\left\langle \widehat{\delta\Delta x_{n+1}}\left(\overline{\widehat{\Xi_{n+1}}}-\overline{\widehat{\Xi_{n}}}\right)\right\rangle \right]$, can be simplified by multiplying Eq.~\ref{eq:linear} by $\overline{\Xi_n}$, averaging, and neglecting the  effect of the noise affecting one car $k$ on the \emph{car in front} of it, $k+1$, viz. $\forall k,\ \left\langle \widehat{\delta x_{k+1}}\ \overline{\widehat{\Xi_{k}}}\right\rangle \approx0$,  so that (writing $n=k+1$)
\begin{equation}
    \forall k,\ -\left(\mathcal{N}+\mathcal{P}\right)\left\langle \widehat{\delta\Delta x_{k+1}}\ \overline{\widehat{\Xi_{k+1}}}\right\rangle \approx \Vert\widehat{\Xi_{k+1}}\Vert^{2} 
\end{equation}
and 
\begin{equation}
\mathrm{CrCor} \approx -2\Vert\widehat{\Xi_{n+1}}\Vert^{2}\mathcal{R}e\left(\frac{\mathcal{P}}{\mathcal{N}+\mathcal{P}}   \right).
\end{equation}

Summing over n and dropping the superfluous $n$ subscripts, we arrive at 
\begin{eqnarray}
    \left\Vert \widehat{\delta\Delta x}\right\Vert ^{2}	&=&	2\frac{1-\mathcal{R}e\left(\frac{\mathcal{P}}{\mathcal{N}+\mathcal{P}}\right)}{|\mathcal{N}+\mathcal{P}|^{2}-\left|\mathcal{P}\right|^{2}} \Vert \widehat{\Xi}\Vert ^{2}\\
	&=&	\overset{f(\omega,\lambda,T)}
    {
    \overbrace{
    \frac{2}{\omega^{4}+\omega^{2}\,(\lambda^{2}+T^{-2})+\lambda^{2}T^{-2}}
    }
    }
    \Vert \widehat{\Xi}\Vert ^{2}.
\end{eqnarray}

Incidentally, we observe that low-frequency vibrations ($\omega \to 0$) are equally weighted, whereas high frequencies are damped because of the finite response time.

Now, we make use of Parseval's theorem to come back to real time space,
\begin{eqnarray}
\langle (\delta \Delta x)^{2}(t)\rangle &=&(2 \pi T_{sim})^{-1} \int_{-\infty}^{\infty}\Vert \widehat{\delta\Delta x}(\omega)\Vert ^2 d\omega\\
&=&(2 \pi T_{sim})^{-1} \int_{-\infty}^{\infty} f(\omega,\lambda,T) \Vert \widehat{\Xi}\Vert ^{2} d\omega.
\end{eqnarray}

Finally, equating the mean-square gap fluctuations $\langle (\delta \Delta x)^{2}(t)\rangle$ obtained with white noise of volatility $\sigma$ (in which case $\Vert \widehat{\Xi}\Vert ^{2}=\sigma^2 T_{sim}$) with those obtained under periodic driving (in which case
\begin{equation}
\Xi_n(t)= C\,\cos(\tilde{\omega} t+ \varphi_n)\text{, i.e., }\widehat{\Xi_n}(\omega)=\pi\,C[ e^{i\varphi_n}\,\delta(\omega - \tilde{\omega})+ e^{-i\varphi_n}\,\delta(\omega + \tilde{\omega})],
\end{equation} 
where $\delta(\bullet)$ tends to a Dirac distribution), we arrive at our final result

\begin{equation}
   C^{2}=A^{2}(\lambda,T)\,\sigma^{2}\text{, where }A^{2}(\lambda,T)=\frac{1}{\pi}\int_{-\infty}^{\infty} \frac{f(\omega,\lambda,T)}{f(\tilde{\omega},\lambda,T)} d\omega \simeq \frac{\lambda}{1+T\lambda}.
   \label{eq:A_final}
\end{equation}

In the last line, we assumed the limit $\tilde{\omega}^2 \ll \min(\lambda^2,T^{-2})$ and we made use of the following identity (obtained from \emph{Mathematica}): 
$\int_{-\infty}^{\infty} \frac{1}{\omega^4 + (\lambda^2+T^{-2})\omega^2 + \lambda^2T^{-2}}d\omega = \pi\frac{T}{\lambda\,(\lambda + T^{-1})}$.

\subparagraph*{Comparison of the theoretically derived coefficient with the actual coefficient.}

Let us now compare the theoretically derived coefficients $A(\lambda,T)$ (Eq.~\ref{eq:A_final}) with those obtained numerically as ratios between $\sigma^{\star}$ and the critical amplitude $C^{\star}$ given by our extended stability analysis, under the random phase approximation of Eq.~\ref{eq:nu_eff}. The results are presented in Table~\ref{tab:coeff}.

The first observation is that the agreement is not strictly quantitative; the observed coefficients $A(\lambda,T)$ are typically two or three times larger than their theoretical counterparts. That being said, the theoretical estimates have the correct order of magnitude and, perhaps more importantly, they mirror the main trends observed numerically, namely:
\begin{itemize}
    \item The coefficients do not depend on the density,
    \item They are virtually insensitive to the frequency $\tilde{\omega}$ of the periodic driving, up to a value around $0.1\,\mathrm{rad/s}$ (for $\lambda=0.2\,\mathrm{s^{-1}}$),
    \item For $\lambda=0.2\,\mathrm{s^{-1}}$, they are approximately constant, regardless of the value of $T$
    \item The coefficients get larger for $\lambda=1\,\mathrm{s^{-1}}$ (about twice larger in the observations, somewhat less in the theoretical predictions), especially for $T=0.5$ s.
\end{itemize}
\begin{table}[h]
\begin{centering}
\begin{tabular}{|c|c|c|c|}
\hline 
 & \cellcolor{yellow!5} T=0.5 & \cellcolor{yellow!5} T=1 & \cellcolor{yellow!5} T=2\tabularnewline
\hline 
\hline 
\multicolumn{4}{c}{Observed values}\tabularnewline
\hline \cellcolor{yellow!5} 
$\lambda=0.2$ & $\approx 1.1a_n$  & \cellcolor{blue!5} $a_n\hat{=}1.0$ & $\simeq a_n$\tabularnewline
\hline \cellcolor{yellow!5} 
$\lambda=1$ & $\approx 2.7a_n$ & $\simeq 1.9a_n$ & \tabularnewline
\hline 
\hline 
\multicolumn{4}{c}{Theoretical values (Eq.~\ref{eq:A_final})}\tabularnewline
\hline 
\cellcolor{yellow!5} $\lambda=0.2$ & $1.05\,a_t$ & \cellcolor{blue!5} $a_t\hat{=}0.41$ & $0.93\,a_t$\tabularnewline
\hline 
\cellcolor{yellow!5} $\lambda=1$ & $2.0\,a_t$ & $1.73\,a_t$ & \tabularnewline
\hline 

\end{tabular}
\par\end{centering}
\caption{Actual values of the prefactor $A(\lambda,T)$. To highlight the variations with $\lambda$, $T$, we present the values of $A(\lambda,T)$ relative to $a_n,\,a_t=A(\lambda=0.2,T=1)$, the values observed numerically or found theoretically for $\lambda=0.2$ and $T=1$.}
\label{tab:coeff}
\end{table}


\begin{thebibliography}{62}%
\makeatletter
\providecommand \@ifxundefined [1]{%
 \@ifx{#1\undefined}
}%
\providecommand \@ifnum [1]{%
 \ifnum #1\expandafter \@firstoftwo
 \else \expandafter \@secondoftwo
 \fi
}%
\providecommand \@ifx [1]{%
 \ifx #1\expandafter \@firstoftwo
 \else \expandafter \@secondoftwo
 \fi
}%
\providecommand \natexlab [1]{#1}%
\providecommand \enquote  [1]{``#1''}%
\providecommand \bibnamefont  [1]{#1}%
\providecommand \bibfnamefont [1]{#1}%
\providecommand \citenamefont [1]{#1}%
\providecommand \href@noop [0]{\@secondoftwo}%
\providecommand \href [0]{\begingroup \@sanitize@url \@href}%
\providecommand \@href[1]{\@@startlink{#1}\@@href}%
\providecommand \@@href[1]{\endgroup#1\@@endlink}%
\providecommand \@sanitize@url [0]{\catcode `\\12\catcode `\$12\catcode
  `\&12\catcode `\#12\catcode `\^12\catcode `\_12\catcode `\%12\relax}%
\providecommand \@@startlink[1]{}%
\providecommand \@@endlink[0]{}%
\providecommand \url  [0]{\begingroup\@sanitize@url \@url }%
\providecommand \@url [1]{\endgroup\@href {#1}{\urlprefix }}%
\providecommand \urlprefix  [0]{URL }%
\providecommand \Eprint [0]{\href }%
\providecommand \doibase [0]{https://doi.org/}%
\providecommand \selectlanguage [0]{\@gobble}%
\providecommand \bibinfo  [0]{\@secondoftwo}%
\providecommand \bibfield  [0]{\@secondoftwo}%
\providecommand \translation [1]{[#1]}%
\providecommand \BibitemOpen [0]{}%
\providecommand \bibitemStop [0]{}%
\providecommand \bibitemNoStop [0]{.\EOS\space}%
\providecommand \EOS [0]{\spacefactor3000\relax}%
\providecommand \BibitemShut  [1]{\csname bibitem#1\endcsname}%
\let\auto@bib@innerbib\@empty
\bibitem [{\citenamefont {Gardiner}(2009)}]{gardiner2009stochastic}%
  \BibitemOpen
  \bibfield  {author} {\bibinfo {author} {\bibfnamefont {C.}~\bibnamefont
  {Gardiner}},\ }\href@noop {} {\emph {\bibinfo {title} {Stochastic
  methods}}},\ Vol.~\bibinfo {volume} {4}\ (\bibinfo  {publisher} {Springer
  Berlin},\ \bibinfo {year} {2009})\BibitemShut {NoStop}%
\bibitem [{\citenamefont {Kerner}(2023)}]{Kerner23}%
  \BibitemOpen
  \bibfield  {author} {\bibinfo {author} {\bibfnamefont {B.}~\bibnamefont
  {Kerner}},\ }\bibfield  {title} {\bibinfo {title} {Model of driver
  overacceleration causing breakdown in vehicular traffic},\ }\href
  {https://doi.org/10.1103/PhysRevE.108.064305} {\bibfield  {journal} {\bibinfo
   {journal} {Phys. Rev. E}\ }\textbf {\bibinfo {volume} {108}},\ \bibinfo
  {pages} {064305} (\bibinfo {year} {2023})}\BibitemShut {NoStop}%
\bibitem [{\citenamefont {Stern}\ \emph {et~al.}(2018)\citenamefont {Stern},
  \citenamefont {Cui}, \citenamefont {Delle~Monache}, \citenamefont {Bhadani},
  \citenamefont {Bunting}, \citenamefont {Churchill}, \citenamefont {Hamilton},
  \citenamefont {Haulcy}, \citenamefont {Pohlmann}, \citenamefont {Wu},
  \citenamefont {Piccoli}, \citenamefont {Seibold}, \citenamefont {Sprinkle},\
  and\ \citenamefont {Work}}]{stern_DissipationStopandgoWaves_2018}%
  \BibitemOpen
  \bibfield  {author} {\bibinfo {author} {\bibfnamefont {R.~E.}\ \bibnamefont
  {Stern}}, \bibinfo {author} {\bibfnamefont {S.}~\bibnamefont {Cui}}, \bibinfo
  {author} {\bibfnamefont {M.~L.}\ \bibnamefont {Delle~Monache}}, \bibinfo
  {author} {\bibfnamefont {R.}~\bibnamefont {Bhadani}}, \bibinfo {author}
  {\bibfnamefont {M.}~\bibnamefont {Bunting}}, \bibinfo {author} {\bibfnamefont
  {M.}~\bibnamefont {Churchill}}, \bibinfo {author} {\bibfnamefont
  {N.}~\bibnamefont {Hamilton}}, \bibinfo {author} {\bibfnamefont
  {R.}~\bibnamefont {Haulcy}}, \bibinfo {author} {\bibfnamefont
  {H.}~\bibnamefont {Pohlmann}}, \bibinfo {author} {\bibfnamefont
  {F.}~\bibnamefont {Wu}}, \bibinfo {author} {\bibfnamefont {B.}~\bibnamefont
  {Piccoli}}, \bibinfo {author} {\bibfnamefont {B.}~\bibnamefont {Seibold}},
  \bibinfo {author} {\bibfnamefont {J.}~\bibnamefont {Sprinkle}},\ and\
  \bibinfo {author} {\bibfnamefont {D.~B.}\ \bibnamefont {Work}},\ }\bibfield
  {title} {\bibinfo {title} {Dissipation of stop-and-go waves via control of
  autonomous vehicles: {{Field}} experiments},\ }\href
  {https://doi.org/10.1016/j.trc.2018.02.005} {\bibfield  {journal} {\bibinfo
  {journal} {Transportation Research Part C: Emerging Technologies}\ }\textbf
  {\bibinfo {volume} {89}},\ \bibinfo {pages} {205} (\bibinfo {year}
  {2018})}\BibitemShut {NoStop}%
\bibitem [{\citenamefont {Gunter}\ \emph {et~al.}(2020)\citenamefont {Gunter},
  \citenamefont {Gloudemans}, \citenamefont {Stern}, \citenamefont {McQuade},
  \citenamefont {Bhadani}, \citenamefont {Bunting}, \citenamefont
  {Delle~Monache}, \citenamefont {Lysecky}, \citenamefont {Seibold},
  \citenamefont {Sprinkle} \emph {et~al.}}]{gunter2020commercially}%
  \BibitemOpen
  \bibfield  {author} {\bibinfo {author} {\bibfnamefont {G.}~\bibnamefont
  {Gunter}}, \bibinfo {author} {\bibfnamefont {D.}~\bibnamefont {Gloudemans}},
  \bibinfo {author} {\bibfnamefont {R.~E.}\ \bibnamefont {Stern}}, \bibinfo
  {author} {\bibfnamefont {S.}~\bibnamefont {McQuade}}, \bibinfo {author}
  {\bibfnamefont {R.}~\bibnamefont {Bhadani}}, \bibinfo {author} {\bibfnamefont
  {M.}~\bibnamefont {Bunting}}, \bibinfo {author} {\bibfnamefont {M.~L.}\
  \bibnamefont {Delle~Monache}}, \bibinfo {author} {\bibfnamefont
  {R.}~\bibnamefont {Lysecky}}, \bibinfo {author} {\bibfnamefont
  {B.}~\bibnamefont {Seibold}}, \bibinfo {author} {\bibfnamefont
  {J.}~\bibnamefont {Sprinkle}}, \emph {et~al.},\ }\bibfield  {title} {\bibinfo
  {title} {Are commercially implemented adaptive cruise control systems string
  stable?},\ }\href@noop {} {\bibfield  {journal} {\bibinfo  {journal} {IEEE
  Transactions on Intelligent Transportation Systems}\ }\textbf {\bibinfo
  {volume} {22}},\ \bibinfo {pages} {6992} (\bibinfo {year}
  {2020})}\BibitemShut {NoStop}%
\bibitem [{\citenamefont {Makridis}\ \emph {et~al.}(2021)\citenamefont
  {Makridis}, \citenamefont {Mattas}, \citenamefont {Anesiadou},\ and\
  \citenamefont {Ciuffo}}]{makridis2021openacc}%
  \BibitemOpen
  \bibfield  {author} {\bibinfo {author} {\bibfnamefont {M.}~\bibnamefont
  {Makridis}}, \bibinfo {author} {\bibfnamefont {K.}~\bibnamefont {Mattas}},
  \bibinfo {author} {\bibfnamefont {A.}~\bibnamefont {Anesiadou}},\ and\
  \bibinfo {author} {\bibfnamefont {B.}~\bibnamefont {Ciuffo}},\ }\bibfield
  {title} {\bibinfo {title} {{OpenACC}. an open database of car-following
  experiments to study the properties of commercial {ACC} systems},\
  }\href@noop {} {\bibfield  {journal} {\bibinfo  {journal} {Transportation
  Research Part C: Emerging Technologies}\ }\textbf {\bibinfo {volume} {125}},\
  \bibinfo {pages} {103047} (\bibinfo {year} {2021})}\BibitemShut {NoStop}%
\bibitem [{\citenamefont {Nagatani}\ and\ \citenamefont
  {Nakanishi}(1998)}]{nagatani1998delay}%
  \BibitemOpen
  \bibfield  {author} {\bibinfo {author} {\bibfnamefont {T.}~\bibnamefont
  {Nagatani}}\ and\ \bibinfo {author} {\bibfnamefont {K.}~\bibnamefont
  {Nakanishi}},\ }\bibfield  {title} {\bibinfo {title} {Delay effect on phase
  transitions in traffic dynamics},\ }\href@noop {} {\bibfield  {journal}
  {\bibinfo  {journal} {Physical Review E}\ }\textbf {\bibinfo {volume} {57}},\
  \bibinfo {pages} {6415} (\bibinfo {year} {1998})}\BibitemShut {NoStop}%
\bibitem [{\citenamefont {Yeo}\ and\ \citenamefont
  {Skabardonis}(2009)}]{yeo2009understanding}%
  \BibitemOpen
  \bibfield  {author} {\bibinfo {author} {\bibfnamefont {H.}~\bibnamefont
  {Yeo}}\ and\ \bibinfo {author} {\bibfnamefont {A.}~\bibnamefont
  {Skabardonis}},\ }\bibfield  {title} {\bibinfo {title} {Understanding
  stop-and-go traffic in view of asymmetric traffic theory},\ }in\ \href@noop
  {} {\emph {\bibinfo {booktitle} {Transportation and Traffic Theory 2009:
  Golden Jubilee: Papers selected for presentation at ISTTT18, a peer reviewed
  series since 1959}}}\ (\bibinfo  {publisher} {Springer},\ \bibinfo {year}
  {2009})\ pp.\ \bibinfo {pages} {99--115}\BibitemShut {NoStop}%
\bibitem [{\citenamefont {Laval}\ \emph {et~al.}(2014)\citenamefont {Laval},
  \citenamefont {Toth},\ and\ \citenamefont {Zhou}}]{laval2014parsimonious}%
  \BibitemOpen
  \bibfield  {author} {\bibinfo {author} {\bibfnamefont {J.~A.}\ \bibnamefont
  {Laval}}, \bibinfo {author} {\bibfnamefont {C.~S.}\ \bibnamefont {Toth}},\
  and\ \bibinfo {author} {\bibfnamefont {Y.}~\bibnamefont {Zhou}},\ }\bibfield
  {title} {\bibinfo {title} {A parsimonious model for the formation of
  oscillations in car-following models},\ }\href@noop {} {\bibfield  {journal}
  {\bibinfo  {journal} {Transportation Research Part B: Methodological}\
  }\textbf {\bibinfo {volume} {70}},\ \bibinfo {pages} {228} (\bibinfo {year}
  {2014})}\BibitemShut {NoStop}%
\bibitem [{\citenamefont {Orosz}\ \emph
  {et~al.}(2004{\natexlab{a}})\citenamefont {Orosz}, \citenamefont {Wilson},\
  and\ \citenamefont {Krauskopf}}]{orosz2004global}%
  \BibitemOpen
  \bibfield  {author} {\bibinfo {author} {\bibfnamefont {G.}~\bibnamefont
  {Orosz}}, \bibinfo {author} {\bibfnamefont {R.}~\bibnamefont {Wilson}},\ and\
  \bibinfo {author} {\bibfnamefont {B.}~\bibnamefont {Krauskopf}},\ }\bibfield
  {title} {\bibinfo {title} {Global bifurcation investigation of an optimal
  velocity traffic model with driver reaction time},\ }\href@noop {} {\bibfield
   {journal} {\bibinfo  {journal} {Physical Review E}\ }\textbf {\bibinfo
  {volume} {70}},\ \bibinfo {pages} {026207} (\bibinfo {year}
  {2004}{\natexlab{a}})}\BibitemShut {NoStop}%
\bibitem [{\citenamefont {Orosz}\ \emph {et~al.}(2010)\citenamefont {Orosz},
  \citenamefont {Wilson},\ and\ \citenamefont
  {St{\'e}p{\'a}n}}]{orosz2010traffic}%
  \BibitemOpen
  \bibfield  {author} {\bibinfo {author} {\bibfnamefont {G.}~\bibnamefont
  {Orosz}}, \bibinfo {author} {\bibfnamefont {R.}~\bibnamefont {Wilson}},\ and\
  \bibinfo {author} {\bibfnamefont {G.}~\bibnamefont {St{\'e}p{\'a}n}},\
  }\bibfield  {title} {\bibinfo {title} {Traffic jams: dynamics and control},\
  }\href@noop {} {\bibfield  {journal} {\bibinfo  {journal} {Philosophical
  Transactions of the Royal Society A: Mathematical, Physical and Engineering
  Sciences}\ }\textbf {\bibinfo {volume} {368}},\ \bibinfo {pages} {4455}
  (\bibinfo {year} {2010})}\BibitemShut {NoStop}%
\bibitem [{\citenamefont {Wilson}\ and\ \citenamefont
  {Ward}(2011)}]{wilson_CarfollowingModelsFifty_2011}%
  \BibitemOpen
  \bibfield  {author} {\bibinfo {author} {\bibfnamefont {R.}~\bibnamefont
  {Wilson}}\ and\ \bibinfo {author} {\bibfnamefont {J.}~\bibnamefont {Ward}},\
  }\bibfield  {title} {\bibinfo {title} {Car-following models: Fifty years of
  linear stability analysis \textendash{} a mathematical perspective},\ }\href
  {https://doi.org/10.1080/03081060.2011.530826} {\bibfield  {journal}
  {\bibinfo  {journal} {Transportation Planning and Technology}\ }\textbf
  {\bibinfo {volume} {34}},\ \bibinfo {pages} {3} (\bibinfo {year}
  {2011})}\BibitemShut {NoStop}%
\bibitem [{\citenamefont {Tordeux}\ \emph {et~al.}(2012)\citenamefont
  {Tordeux}, \citenamefont {Roussignol},\ and\ \citenamefont
  {Lassarre}}]{tordeux_LinearStabilityAnalysis_2012}%
  \BibitemOpen
  \bibfield  {author} {\bibinfo {author} {\bibfnamefont {A.}~\bibnamefont
  {Tordeux}}, \bibinfo {author} {\bibfnamefont {M.}~\bibnamefont
  {Roussignol}},\ and\ \bibinfo {author} {\bibfnamefont {S.}~\bibnamefont
  {Lassarre}},\ }\bibfield  {title} {\bibinfo {title} {Linear stability
  analysis of first-order delayed car-following models on a ring},\ }\href
  {https://doi.org/10.1103/PhysRevE.86.036207} {\bibfield  {journal} {\bibinfo
  {journal} {Physical Review E}\ }\textbf {\bibinfo {volume} {86}},\ \bibinfo
  {pages} {036207} (\bibinfo {year} {2012})}\BibitemShut {NoStop}%
\bibitem [{\citenamefont {Tordeux}\ \emph {et~al.}(2018)\citenamefont
  {Tordeux}, \citenamefont {Costeseque}, \citenamefont {Herty},\ and\
  \citenamefont {Seyfried}}]{tordeux2018traffic}%
  \BibitemOpen
  \bibfield  {author} {\bibinfo {author} {\bibfnamefont {A.}~\bibnamefont
  {Tordeux}}, \bibinfo {author} {\bibfnamefont {G.}~\bibnamefont {Costeseque}},
  \bibinfo {author} {\bibfnamefont {M.}~\bibnamefont {Herty}},\ and\ \bibinfo
  {author} {\bibfnamefont {A.}~\bibnamefont {Seyfried}},\ }\bibfield  {title}
  {\bibinfo {title} {From traffic and pedestrian follow-the-leader models with
  reaction time to first order convection-diffusion flow models},\ }\href@noop
  {} {\bibfield  {journal} {\bibinfo  {journal} {SIAM Journal on Applied
  Mathematics}\ }\textbf {\bibinfo {volume} {78}},\ \bibinfo {pages} {63}
  (\bibinfo {year} {2018})}\BibitemShut {NoStop}%
\bibitem [{\citenamefont {Komatsu}\ and\ \citenamefont
  {Sasa}(1995)}]{komatsu1995kink}%
  \BibitemOpen
  \bibfield  {author} {\bibinfo {author} {\bibfnamefont {T.~S.}\ \bibnamefont
  {Komatsu}}\ and\ \bibinfo {author} {\bibfnamefont {S.-i.}\ \bibnamefont
  {Sasa}},\ }\bibfield  {title} {\bibinfo {title} {Kink soliton characterizing
  traffic congestion},\ }\href@noop {} {\bibfield  {journal} {\bibinfo
  {journal} {Physical Review E}\ }\textbf {\bibinfo {volume} {52}},\ \bibinfo
  {pages} {5574} (\bibinfo {year} {1995})}\BibitemShut {NoStop}%
\bibitem [{\citenamefont {Reuschel}(1950)}]{reuschel1950fahrzeugbewegungen}%
  \BibitemOpen
  \bibfield  {author} {\bibinfo {author} {\bibfnamefont {A.}~\bibnamefont
  {Reuschel}},\ }\bibfield  {title} {\bibinfo {title} {Fahrzeugbewegungen in
  der {K}olonne},\ }\href@noop {} {\bibfield  {journal} {\bibinfo  {journal}
  {\"Osterreichisches Ingenieur Archiv}\ }\textbf {\bibinfo {volume} {4}},\
  \bibinfo {pages} {193} (\bibinfo {year} {1950})}\BibitemShut {NoStop}%
\bibitem [{\citenamefont
  {Pipes}(1953)}]{pipes_OperationalAnalysisTraffic_1953}%
  \BibitemOpen
  \bibfield  {author} {\bibinfo {author} {\bibfnamefont {L.~A.}\ \bibnamefont
  {Pipes}},\ }\bibfield  {title} {\bibinfo {title} {An operational analysis of
  traffic dynamics},\ }\href {https://doi.org/10.1063/1.1721265} {\bibfield
  {journal} {\bibinfo  {journal} {Journal of Applied Physics}\ }\textbf
  {\bibinfo {volume} {24}},\ \bibinfo {pages} {274} (\bibinfo {year}
  {1953})}\BibitemShut {NoStop}%
\bibitem [{\citenamefont {Kometani}\ and\ \citenamefont
  {Sasaki}(1958)}]{kometani1958stability}%
  \BibitemOpen
  \bibfield  {author} {\bibinfo {author} {\bibfnamefont {E.}~\bibnamefont
  {Kometani}}\ and\ \bibinfo {author} {\bibfnamefont {T.}~\bibnamefont
  {Sasaki}},\ }\bibfield  {title} {\bibinfo {title} {On the stability of
  traffic flow (report-i)},\ }\href@noop {} {\bibfield  {journal} {\bibinfo
  {journal} {Journal of the Operations Research Society of Japan}\ }\textbf
  {\bibinfo {volume} {2}},\ \bibinfo {pages} {11} (\bibinfo {year}
  {1958})}\BibitemShut {NoStop}%
\bibitem [{\citenamefont {Chandler}\ \emph {et~al.}(1958)\citenamefont
  {Chandler}, \citenamefont {Herman},\ and\ \citenamefont
  {Montroll}}]{chandler1958traffic}%
  \BibitemOpen
  \bibfield  {author} {\bibinfo {author} {\bibfnamefont {R.~E.}\ \bibnamefont
  {Chandler}}, \bibinfo {author} {\bibfnamefont {R.}~\bibnamefont {Herman}},\
  and\ \bibinfo {author} {\bibfnamefont {E.~W.}\ \bibnamefont {Montroll}},\
  }\bibfield  {title} {\bibinfo {title} {Traffic dynamics: studies in car
  following},\ }\href@noop {} {\bibfield  {journal} {\bibinfo  {journal}
  {Operations Research}\ }\textbf {\bibinfo {volume} {6}},\ \bibinfo {pages}
  {165} (\bibinfo {year} {1958})}\BibitemShut {NoStop}%
\bibitem [{\citenamefont {Herman}\ \emph {et~al.}(1959)\citenamefont {Herman},
  \citenamefont {Montroll}, \citenamefont {Potts},\ and\ \citenamefont
  {Rothery}}]{herman_TrafficDynamicsAnalysis_1959}%
  \BibitemOpen
  \bibfield  {author} {\bibinfo {author} {\bibfnamefont {R.}~\bibnamefont
  {Herman}}, \bibinfo {author} {\bibfnamefont {E.~W.}\ \bibnamefont
  {Montroll}}, \bibinfo {author} {\bibfnamefont {R.~B.}\ \bibnamefont
  {Potts}},\ and\ \bibinfo {author} {\bibfnamefont {R.~W.}\ \bibnamefont
  {Rothery}},\ }\bibfield  {title} {\bibinfo {title} {Traffic dynamics:
  {{Analysis}} of stability in car following},\ }\href@noop {} {\bibfield
  {journal} {\bibinfo  {journal} {Operations Research}\ }\textbf {\bibinfo
  {volume} {7}},\ \bibinfo {pages} {86} (\bibinfo {year} {1959})}\BibitemShut
  {NoStop}%
\bibitem [{\citenamefont {Bando}\ \emph {et~al.}(1995)\citenamefont {Bando},
  \citenamefont {Hasebe}, \citenamefont {Nakayama}, \citenamefont {Shibata},\
  and\ \citenamefont {Sugiyama}}]{bando_DynamicalModelTraffic_1995}%
  \BibitemOpen
  \bibfield  {author} {\bibinfo {author} {\bibfnamefont {M.}~\bibnamefont
  {Bando}}, \bibinfo {author} {\bibfnamefont {K.}~\bibnamefont {Hasebe}},
  \bibinfo {author} {\bibfnamefont {A.}~\bibnamefont {Nakayama}}, \bibinfo
  {author} {\bibfnamefont {A.}~\bibnamefont {Shibata}},\ and\ \bibinfo {author}
  {\bibfnamefont {Y.}~\bibnamefont {Sugiyama}},\ }\bibfield  {title} {\bibinfo
  {title} {Dynamical model of traffic congestion and numerical simulation},\
  }\href {https://doi.org/10.1103/PhysRevE.51.1035} {\bibfield  {journal}
  {\bibinfo  {journal} {Physical Review E}\ }\textbf {\bibinfo {volume} {51}},\
  \bibinfo {pages} {1035} (\bibinfo {year} {1995})}\BibitemShut {NoStop}%
\bibitem [{\citenamefont {Bando}\ \emph {et~al.}(1998)\citenamefont {Bando},
  \citenamefont {Hasebe}, \citenamefont {Nakanishi},\ and\ \citenamefont
  {Nakayama}}]{bando1998analysis}%
  \BibitemOpen
  \bibfield  {author} {\bibinfo {author} {\bibfnamefont {M.}~\bibnamefont
  {Bando}}, \bibinfo {author} {\bibfnamefont {K.}~\bibnamefont {Hasebe}},
  \bibinfo {author} {\bibfnamefont {K.}~\bibnamefont {Nakanishi}},\ and\
  \bibinfo {author} {\bibfnamefont {A.}~\bibnamefont {Nakayama}},\ }\bibfield
  {title} {\bibinfo {title} {Analysis of optimal velocity model with explicit
  delay},\ }\href@noop {} {\bibfield  {journal} {\bibinfo  {journal} {Physical
  Review E}\ }\textbf {\bibinfo {volume} {58}},\ \bibinfo {pages} {5429}
  (\bibinfo {year} {1998})}\BibitemShut {NoStop}%
\bibitem [{\citenamefont {Makridis}\ \emph {et~al.}(2019)\citenamefont
  {Makridis}, \citenamefont {Mattas},\ and\ \citenamefont
  {Ciuffo}}]{makridis2019response}%
  \BibitemOpen
  \bibfield  {author} {\bibinfo {author} {\bibfnamefont {M.}~\bibnamefont
  {Makridis}}, \bibinfo {author} {\bibfnamefont {K.}~\bibnamefont {Mattas}},\
  and\ \bibinfo {author} {\bibfnamefont {B.}~\bibnamefont {Ciuffo}},\
  }\bibfield  {title} {\bibinfo {title} {Response time and time headway of an
  adaptive cruise control. an empirical characterization and potential impacts
  on road capacity},\ }\href@noop {} {\bibfield  {journal} {\bibinfo  {journal}
  {IEEE transactions on intelligent transportation systems}\ }\textbf {\bibinfo
  {volume} {21}},\ \bibinfo {pages} {1677} (\bibinfo {year}
  {2019})}\BibitemShut {NoStop}%
\bibitem [{\citenamefont {Treiber}\ \emph {et~al.}(2000)\citenamefont
  {Treiber}, \citenamefont {Hennecke},\ and\ \citenamefont
  {Helbing}}]{treiber_CongestedTrafficStates_2000}%
  \BibitemOpen
  \bibfield  {author} {\bibinfo {author} {\bibfnamefont {M.}~\bibnamefont
  {Treiber}}, \bibinfo {author} {\bibfnamefont {A.}~\bibnamefont {Hennecke}},\
  and\ \bibinfo {author} {\bibfnamefont {D.}~\bibnamefont {Helbing}},\
  }\bibfield  {title} {\bibinfo {title} {Congested traffic states in empirical
  observations and microscopic simulations},\ }\href
  {https://doi.org/10.1103/PhysRevE.62.1805} {\bibfield  {journal} {\bibinfo
  {journal} {Physical Review E}\ }\textbf {\bibinfo {volume} {62}},\ \bibinfo
  {pages} {1805} (\bibinfo {year} {2000})}\BibitemShut {NoStop}%
\bibitem [{\citenamefont {Jiang}\ \emph {et~al.}(2001)\citenamefont {Jiang},
  \citenamefont {Wu},\ and\ \citenamefont
  {Zhu}}]{jiang_FullVelocityDifference_2001}%
  \BibitemOpen
  \bibfield  {author} {\bibinfo {author} {\bibfnamefont {R.}~\bibnamefont
  {Jiang}}, \bibinfo {author} {\bibfnamefont {Q.}~\bibnamefont {Wu}},\ and\
  \bibinfo {author} {\bibfnamefont {Z.}~\bibnamefont {Zhu}},\ }\bibfield
  {title} {\bibinfo {title} {Full velocity difference model for a car-following
  theory},\ }\href {https://doi.org/10.1103/PhysRevE.64.017101} {\bibfield
  {journal} {\bibinfo  {journal} {Physical Review E}\ }\textbf {\bibinfo
  {volume} {64}},\ \bibinfo {pages} {017101} (\bibinfo {year}
  {2001})}\BibitemShut {NoStop}%
\bibitem [{\citenamefont {Tomer}\ \emph {et~al.}(2000)\citenamefont {Tomer},
  \citenamefont {Safonov},\ and\ \citenamefont {Havlin}}]{tomer2000presence}%
  \BibitemOpen
  \bibfield  {author} {\bibinfo {author} {\bibfnamefont {E.}~\bibnamefont
  {Tomer}}, \bibinfo {author} {\bibfnamefont {L.}~\bibnamefont {Safonov}},\
  and\ \bibinfo {author} {\bibfnamefont {S.}~\bibnamefont {Havlin}},\
  }\bibfield  {title} {\bibinfo {title} {Presence of many stable nonhomogeneous
  states in an inertial car-following model},\ }\href@noop {} {\bibfield
  {journal} {\bibinfo  {journal} {Physical Review Letters}\ }\textbf {\bibinfo
  {volume} {84}},\ \bibinfo {pages} {382} (\bibinfo {year} {2000})}\BibitemShut
  {NoStop}%
\bibitem [{\citenamefont {Tordeux}\ \emph {et~al.}(2010)\citenamefont
  {Tordeux}, \citenamefont {Lassarre},\ and\ \citenamefont
  {Roussignol}}]{tordeux2010adaptive}%
  \BibitemOpen
  \bibfield  {author} {\bibinfo {author} {\bibfnamefont {A.}~\bibnamefont
  {Tordeux}}, \bibinfo {author} {\bibfnamefont {S.}~\bibnamefont {Lassarre}},\
  and\ \bibinfo {author} {\bibfnamefont {M.}~\bibnamefont {Roussignol}},\
  }\bibfield  {title} {\bibinfo {title} {An adaptive time gap car-following
  model},\ }\href@noop {} {\bibfield  {journal} {\bibinfo  {journal}
  {Transportation Research Part B: Methodological}\ }\textbf {\bibinfo {volume}
  {44}},\ \bibinfo {pages} {1115} (\bibinfo {year} {2010})}\BibitemShut
  {NoStop}%
\bibitem [{\citenamefont {Jiang}\ \emph {et~al.}(2018)\citenamefont {Jiang},
  \citenamefont {Jin}, \citenamefont {Zhang}, \citenamefont {Huang},
  \citenamefont {Tian}, \citenamefont {Wang}, \citenamefont {Hu}, \citenamefont
  {Wang},\ and\ \citenamefont {Jia}}]{jiang2018experimental}%
  \BibitemOpen
  \bibfield  {author} {\bibinfo {author} {\bibfnamefont {R.}~\bibnamefont
  {Jiang}}, \bibinfo {author} {\bibfnamefont {C.-J.}\ \bibnamefont {Jin}},
  \bibinfo {author} {\bibfnamefont {H.}~\bibnamefont {Zhang}}, \bibinfo
  {author} {\bibfnamefont {Y.-X.}\ \bibnamefont {Huang}}, \bibinfo {author}
  {\bibfnamefont {J.-F.}\ \bibnamefont {Tian}}, \bibinfo {author}
  {\bibfnamefont {W.}~\bibnamefont {Wang}}, \bibinfo {author} {\bibfnamefont
  {M.-B.}\ \bibnamefont {Hu}}, \bibinfo {author} {\bibfnamefont
  {H.}~\bibnamefont {Wang}},\ and\ \bibinfo {author} {\bibfnamefont
  {B.}~\bibnamefont {Jia}},\ }\bibfield  {title} {\bibinfo {title}
  {Experimental and empirical investigations of traffic flow instability},\
  }\href@noop {} {\bibfield  {journal} {\bibinfo  {journal} {Transportation
  Research Part C: Rmerging Technologies}\ }\textbf {\bibinfo {volume} {94}},\
  \bibinfo {pages} {83} (\bibinfo {year} {2018})}\BibitemShut {NoStop}%
\bibitem [{\citenamefont {Tian}\ \emph {et~al.}(2016)\citenamefont {Tian},
  \citenamefont {Jiang}, \citenamefont {Jia}, \citenamefont {Gao},\ and\
  \citenamefont {Ma}}]{tian2016empirical}%
  \BibitemOpen
  \bibfield  {author} {\bibinfo {author} {\bibfnamefont {J.}~\bibnamefont
  {Tian}}, \bibinfo {author} {\bibfnamefont {R.}~\bibnamefont {Jiang}},
  \bibinfo {author} {\bibfnamefont {B.}~\bibnamefont {Jia}}, \bibinfo {author}
  {\bibfnamefont {Z.}~\bibnamefont {Gao}},\ and\ \bibinfo {author}
  {\bibfnamefont {S.}~\bibnamefont {Ma}},\ }\bibfield  {title} {\bibinfo
  {title} {Empirical analysis and simulation of the concave growth pattern of
  traffic oscillations},\ }\href@noop {} {\bibfield  {journal} {\bibinfo
  {journal} {Transportation Research Part B: Methodological}\ }\textbf
  {\bibinfo {volume} {93}},\ \bibinfo {pages} {338} (\bibinfo {year}
  {2016})}\BibitemShut {NoStop}%
\bibitem [{\citenamefont {Jiang}\ \emph {et~al.}(2015)\citenamefont {Jiang},
  \citenamefont {Hu}, \citenamefont {Zhang}, \citenamefont {Gao}, \citenamefont
  {Jia},\ and\ \citenamefont {Wu}}]{jiang2015some}%
  \BibitemOpen
  \bibfield  {author} {\bibinfo {author} {\bibfnamefont {R.}~\bibnamefont
  {Jiang}}, \bibinfo {author} {\bibfnamefont {M.-B.}\ \bibnamefont {Hu}},
  \bibinfo {author} {\bibfnamefont {H.}~\bibnamefont {Zhang}}, \bibinfo
  {author} {\bibfnamefont {Z.-Y.}\ \bibnamefont {Gao}}, \bibinfo {author}
  {\bibfnamefont {B.}~\bibnamefont {Jia}},\ and\ \bibinfo {author}
  {\bibfnamefont {Q.-S.}\ \bibnamefont {Wu}},\ }\bibfield  {title} {\bibinfo
  {title} {On some experimental features of car-following behavior and how to
  model them},\ }\href@noop {} {\bibfield  {journal} {\bibinfo  {journal}
  {Transportation Research Part B: Methodological}\ }\textbf {\bibinfo {volume}
  {80}},\ \bibinfo {pages} {338} (\bibinfo {year} {2015})}\BibitemShut
  {NoStop}%
\bibitem [{\citenamefont {Tian}\ \emph {et~al.}(2019)\citenamefont {Tian},
  \citenamefont {Zhang}, \citenamefont {Treiber}, \citenamefont {Jiang},
  \citenamefont {Gao},\ and\ \citenamefont {Jia}}]{tian2019role}%
  \BibitemOpen
  \bibfield  {author} {\bibinfo {author} {\bibfnamefont {J.}~\bibnamefont
  {Tian}}, \bibinfo {author} {\bibfnamefont {H.}~\bibnamefont {Zhang}},
  \bibinfo {author} {\bibfnamefont {M.}~\bibnamefont {Treiber}}, \bibinfo
  {author} {\bibfnamefont {R.}~\bibnamefont {Jiang}}, \bibinfo {author}
  {\bibfnamefont {Z.-Y.}\ \bibnamefont {Gao}},\ and\ \bibinfo {author}
  {\bibfnamefont {B.}~\bibnamefont {Jia}},\ }\bibfield  {title} {\bibinfo
  {title} {On the role of speed adaptation and spacing indifference in traffic
  instability: Evidence from car-following experiments and its stochastic
  model},\ }\href@noop {} {\bibfield  {journal} {\bibinfo  {journal}
  {Transportation Research Part B: Methodological}\ }\textbf {\bibinfo {volume}
  {129}},\ \bibinfo {pages} {334} (\bibinfo {year} {2019})}\BibitemShut
  {NoStop}%
\bibitem [{\citenamefont {Sugiyama}\ \emph {et~al.}(2008)\citenamefont
  {Sugiyama}, \citenamefont {Fukui}, \citenamefont {Kikuchi}, \citenamefont
  {Hasebe}, \citenamefont {Nakayama}, \citenamefont {Nishinari}, \citenamefont
  {Tadaki},\ and\ \citenamefont {Yukawa}}]{sugiyama2008traffic}%
  \BibitemOpen
  \bibfield  {author} {\bibinfo {author} {\bibfnamefont {Y.}~\bibnamefont
  {Sugiyama}}, \bibinfo {author} {\bibfnamefont {M.}~\bibnamefont {Fukui}},
  \bibinfo {author} {\bibfnamefont {M.}~\bibnamefont {Kikuchi}}, \bibinfo
  {author} {\bibfnamefont {K.}~\bibnamefont {Hasebe}}, \bibinfo {author}
  {\bibfnamefont {A.}~\bibnamefont {Nakayama}}, \bibinfo {author}
  {\bibfnamefont {K.}~\bibnamefont {Nishinari}}, \bibinfo {author}
  {\bibfnamefont {S.-i.}\ \bibnamefont {Tadaki}},\ and\ \bibinfo {author}
  {\bibfnamefont {S.}~\bibnamefont {Yukawa}},\ }\bibfield  {title} {\bibinfo
  {title} {Traffic jams without bottlenecks—experimental evidence for the
  physical mechanism of the formation of a jam},\ }\href@noop {} {\bibfield
  {journal} {\bibinfo  {journal} {New Journal of Physics.}\ }\textbf {\bibinfo
  {volume} {10}},\ \bibinfo {pages} {033001} (\bibinfo {year}
  {2008})}\BibitemShut {NoStop}%
\bibitem [{\citenamefont {Tadaki}\ \emph {et~al.}(2013)\citenamefont {Tadaki},
  \citenamefont {Kikuchi}, \citenamefont {Fukui}, \citenamefont {Nakayama},
  \citenamefont {Nishinari}, \citenamefont {Shibata}, \citenamefont {Sugiyama},
  \citenamefont {Yosida},\ and\ \citenamefont {Yukawa}}]{tadaki2013phase}%
  \BibitemOpen
  \bibfield  {author} {\bibinfo {author} {\bibfnamefont {S.-i.}\ \bibnamefont
  {Tadaki}}, \bibinfo {author} {\bibfnamefont {M.}~\bibnamefont {Kikuchi}},
  \bibinfo {author} {\bibfnamefont {M.}~\bibnamefont {Fukui}}, \bibinfo
  {author} {\bibfnamefont {A.}~\bibnamefont {Nakayama}}, \bibinfo {author}
  {\bibfnamefont {K.}~\bibnamefont {Nishinari}}, \bibinfo {author}
  {\bibfnamefont {A.}~\bibnamefont {Shibata}}, \bibinfo {author} {\bibfnamefont
  {Y.}~\bibnamefont {Sugiyama}}, \bibinfo {author} {\bibfnamefont
  {T.}~\bibnamefont {Yosida}},\ and\ \bibinfo {author} {\bibfnamefont
  {S.}~\bibnamefont {Yukawa}},\ }\bibfield  {title} {\bibinfo {title} {Phase
  transition in traffic jam experiment on a circuit},\ }\href@noop {}
  {\bibfield  {journal} {\bibinfo  {journal} {New Journal of Physics}\ }\textbf
  {\bibinfo {volume} {15}},\ \bibinfo {pages} {103034} (\bibinfo {year}
  {2013})}\BibitemShut {NoStop}%
\bibitem [{\citenamefont {Nakayama}\ \emph {et~al.}(2009)\citenamefont
  {Nakayama}, \citenamefont {Fukui}, \citenamefont {Kikuchi}, \citenamefont
  {Hasebe}, \citenamefont {Nishinari}, \citenamefont {Sugiyama}, \citenamefont
  {Tadaki},\ and\ \citenamefont {Yukawa}}]{nakayama2009metastability}%
  \BibitemOpen
  \bibfield  {author} {\bibinfo {author} {\bibfnamefont {A.}~\bibnamefont
  {Nakayama}}, \bibinfo {author} {\bibfnamefont {M.}~\bibnamefont {Fukui}},
  \bibinfo {author} {\bibfnamefont {M.}~\bibnamefont {Kikuchi}}, \bibinfo
  {author} {\bibfnamefont {K.}~\bibnamefont {Hasebe}}, \bibinfo {author}
  {\bibfnamefont {K.}~\bibnamefont {Nishinari}}, \bibinfo {author}
  {\bibfnamefont {Y.}~\bibnamefont {Sugiyama}}, \bibinfo {author}
  {\bibfnamefont {S.-i.}\ \bibnamefont {Tadaki}},\ and\ \bibinfo {author}
  {\bibfnamefont {S.}~\bibnamefont {Yukawa}},\ }\bibfield  {title} {\bibinfo
  {title} {Metastability in the formation of an experimental traffic jam},\
  }\href@noop {} {\bibfield  {journal} {\bibinfo  {journal} {New Journal of
  Physics}\ }\textbf {\bibinfo {volume} {11}},\ \bibinfo {pages} {083025}
  (\bibinfo {year} {2009})}\BibitemShut {NoStop}%
\bibitem [{\citenamefont {Schadschneider}\ \emph {et~al.}(2010)\citenamefont
  {Schadschneider}, \citenamefont {Chowdhury},\ and\ \citenamefont
  {Nishinari}}]{schadschneider_StochasticTransportComplex_2010}%
  \BibitemOpen
  \bibfield  {author} {\bibinfo {author} {\bibfnamefont {A.}~\bibnamefont
  {Schadschneider}}, \bibinfo {author} {\bibfnamefont {D.}~\bibnamefont
  {Chowdhury}},\ and\ \bibinfo {author} {\bibfnamefont {K.}~\bibnamefont
  {Nishinari}},\ }\href@noop {} {\emph {\bibinfo {title} {Stochastic
  {{Transport}} in {{Complex Systems}}. {{From Molecules}} to {{Vehicles}}}}}\
  (\bibinfo  {publisher} {Elsevier},\ \bibinfo {year} {2010})\BibitemShut
  {NoStop}%
\bibitem [{\citenamefont {Treiber}\ and\ \citenamefont
  {Kesting}(2017)}]{treiber2017intelligent}%
  \BibitemOpen
  \bibfield  {author} {\bibinfo {author} {\bibfnamefont {M.}~\bibnamefont
  {Treiber}}\ and\ \bibinfo {author} {\bibfnamefont {A.}~\bibnamefont
  {Kesting}},\ }\bibfield  {title} {\bibinfo {title} {The intelligent driver
  model with stochasticity-new insights into traffic flow oscillations},\
  }\href@noop {} {\bibfield  {journal} {\bibinfo  {journal} {Transportation
  Research Procedia}\ }\textbf {\bibinfo {volume} {23}},\ \bibinfo {pages}
  {174} (\bibinfo {year} {2017})}\BibitemShut {NoStop}%
\bibitem [{\citenamefont {Wiesenfeld}(1985)}]{wiesenfeld1985noisy}%
  \BibitemOpen
  \bibfield  {author} {\bibinfo {author} {\bibfnamefont {K.}~\bibnamefont
  {Wiesenfeld}},\ }\bibfield  {title} {\bibinfo {title} {Noisy precursors of
  nonlinear instabilities},\ }\href@noop {} {\bibfield  {journal} {\bibinfo
  {journal} {Journal of Statistical Physics}\ }\textbf {\bibinfo {volume}
  {38}},\ \bibinfo {pages} {1071} (\bibinfo {year} {1985})}\BibitemShut
  {NoStop}%
\bibitem [{\citenamefont {Ngoduy}(2021)}]{Ngoduy21}%
  \BibitemOpen
  \bibfield  {author} {\bibinfo {author} {\bibfnamefont {D.}~\bibnamefont
  {Ngoduy}},\ }\bibfield  {title} {\bibinfo {title} {Noise-induced instability
  of a class of stochastic higher order continuum traffic models},\ }\href
  {https://doi.org/10.1016/j.trb.2021.06.013} {\bibfield  {journal} {\bibinfo
  {journal} {Transportation Research Part B: Methodological}\ }\textbf
  {\bibinfo {volume} {150}},\ \bibinfo {pages} {260} (\bibinfo {year}
  {2021})}\BibitemShut {NoStop}%
\bibitem [{\citenamefont {Barlovic}\ \emph {et~al.}(1998)\citenamefont
  {Barlovic}, \citenamefont {Santen}, \citenamefont {Schadschneider},\ and\
  \citenamefont {Schreckenberg}}]{BarlovicSSS98}%
  \BibitemOpen
  \bibfield  {author} {\bibinfo {author} {\bibfnamefont {R.}~\bibnamefont
  {Barlovic}}, \bibinfo {author} {\bibfnamefont {L.}~\bibnamefont {Santen}},
  \bibinfo {author} {\bibfnamefont {A.}~\bibnamefont {Schadschneider}},\ and\
  \bibinfo {author} {\bibfnamefont {M.}~\bibnamefont {Schreckenberg}},\
  }\bibfield  {title} {\bibinfo {title} {Metastable states in cellular automata
  for traffic flow},\ }\href
  {https://doi.org/https://doi.org/10.1007/s100510050504} {\bibfield  {journal}
  {\bibinfo  {journal} {Eur. Phys. J. B}\ }\textbf {\bibinfo {volume} {5}},\
  \bibinfo {pages} {793} (\bibinfo {year} {1998})}\BibitemShut {NoStop}%
\bibitem [{\citenamefont {Ke-Ping}\ and\ \citenamefont
  {Zi-You}(2004{\natexlab{a}})}]{Ke-Ping_2004}%
  \BibitemOpen
  \bibfield  {author} {\bibinfo {author} {\bibfnamefont {L.}~\bibnamefont
  {Ke-Ping}}\ and\ \bibinfo {author} {\bibfnamefont {G.}~\bibnamefont
  {Zi-You}},\ }\bibfield  {title} {\bibinfo {title} {Noise-induced phase
  transition in traffic flow*},\ }\href
  {https://doi.org/10.1088/0253-6102/42/3/369} {\bibfield  {journal} {\bibinfo
  {journal} {Communications in Theoretical Physics}\ }\textbf {\bibinfo
  {volume} {42}},\ \bibinfo {pages} {369} (\bibinfo {year}
  {2004}{\natexlab{a}})}\BibitemShut {NoStop}%
\bibitem [{\citenamefont {Kaupu{\v{z}}s}\ \emph {et~al.}(2005)\citenamefont
  {Kaupu{\v{z}}s}, \citenamefont {Mahnke},\ and\ \citenamefont
  {Harris}}]{kaupuvzs2005zero}%
  \BibitemOpen
  \bibfield  {author} {\bibinfo {author} {\bibfnamefont {J.}~\bibnamefont
  {Kaupu{\v{z}}s}}, \bibinfo {author} {\bibfnamefont {R.}~\bibnamefont
  {Mahnke}},\ and\ \bibinfo {author} {\bibfnamefont {R.}~\bibnamefont
  {Harris}},\ }\bibfield  {title} {\bibinfo {title} {Zero-range model of
  traffic flow},\ }\href@noop {} {\bibfield  {journal} {\bibinfo  {journal}
  {Physical Review E}\ }\textbf {\bibinfo {volume} {72}},\ \bibinfo {pages}
  {056125} (\bibinfo {year} {2005})}\BibitemShut {NoStop}%
\bibitem [{\citenamefont {Huang}\ \emph {et~al.}(2018)\citenamefont {Huang},
  \citenamefont {Guo}, \citenamefont {Jiang},\ and\ \citenamefont
  {Hu}}]{huang2018instability}%
  \BibitemOpen
  \bibfield  {author} {\bibinfo {author} {\bibfnamefont {Y.-X.}\ \bibnamefont
  {Huang}}, \bibinfo {author} {\bibfnamefont {N.}~\bibnamefont {Guo}}, \bibinfo
  {author} {\bibfnamefont {R.}~\bibnamefont {Jiang}},\ and\ \bibinfo {author}
  {\bibfnamefont {M.-B.}\ \bibnamefont {Hu}},\ }\bibfield  {title} {\bibinfo
  {title} {Instability in car-following behavior: new nagel--schreckenberg type
  cellular automata model},\ }\href@noop {} {\bibfield  {journal} {\bibinfo
  {journal} {Journal of Statistical Mechanics: Theory and Experiment}\ }\textbf
  {\bibinfo {volume} {2018}},\ \bibinfo {pages} {083401} (\bibinfo {year}
  {2018})}\BibitemShut {NoStop}%
\bibitem [{\citenamefont {Wagner}(2011)}]{wagner2011time}%
  \BibitemOpen
  \bibfield  {author} {\bibinfo {author} {\bibfnamefont {P.}~\bibnamefont
  {Wagner}},\ }\bibfield  {title} {\bibinfo {title} {A time-discrete harmonic
  oscillator model of human car-following},\ }\href@noop {} {\bibfield
  {journal} {\bibinfo  {journal} {The European Physical Journal B}\ }\textbf
  {\bibinfo {volume} {84}},\ \bibinfo {pages} {713} (\bibinfo {year}
  {2011})}\BibitemShut {NoStop}%
\bibitem [{\citenamefont {Treiber}\ and\ \citenamefont
  {Helbing}(2009)}]{treiber2009hamilton}%
  \BibitemOpen
  \bibfield  {author} {\bibinfo {author} {\bibfnamefont {M.}~\bibnamefont
  {Treiber}}\ and\ \bibinfo {author} {\bibfnamefont {D.}~\bibnamefont
  {Helbing}},\ }\bibfield  {title} {\bibinfo {title} {Hamilton-like statistics
  in onedimensional driven dissipative many-particle systems},\ }\href@noop {}
  {\bibfield  {journal} {\bibinfo  {journal} {The European Physical Journal B}\
  }\textbf {\bibinfo {volume} {68}},\ \bibinfo {pages} {607} (\bibinfo {year}
  {2009})}\BibitemShut {NoStop}%
\bibitem [{\citenamefont {Wang}\ \emph {et~al.}(2020)\citenamefont {Wang},
  \citenamefont {Li}, \citenamefont {Tian},\ and\ \citenamefont
  {Jiang}}]{wang2020stability}%
  \BibitemOpen
  \bibfield  {author} {\bibinfo {author} {\bibfnamefont {Y.}~\bibnamefont
  {Wang}}, \bibinfo {author} {\bibfnamefont {X.}~\bibnamefont {Li}}, \bibinfo
  {author} {\bibfnamefont {J.}~\bibnamefont {Tian}},\ and\ \bibinfo {author}
  {\bibfnamefont {R.}~\bibnamefont {Jiang}},\ }\bibfield  {title} {\bibinfo
  {title} {Stability analysis of stochastic linear car-following models},\
  }\href@noop {} {\bibfield  {journal} {\bibinfo  {journal} {Transportation
  Science}\ }\textbf {\bibinfo {volume} {54}},\ \bibinfo {pages} {274}
  (\bibinfo {year} {2020})}\BibitemShut {NoStop}%
\bibitem [{\citenamefont {Friesen}\ \emph {et~al.}(2021)\citenamefont
  {Friesen}, \citenamefont {Gottschalk}, \citenamefont {R{\"u}diger},\ and\
  \citenamefont {Tordeux}}]{friesen2021spontaneous}%
  \BibitemOpen
  \bibfield  {author} {\bibinfo {author} {\bibfnamefont {M.}~\bibnamefont
  {Friesen}}, \bibinfo {author} {\bibfnamefont {H.}~\bibnamefont {Gottschalk}},
  \bibinfo {author} {\bibfnamefont {B.}~\bibnamefont {R{\"u}diger}},\ and\
  \bibinfo {author} {\bibfnamefont {A.}~\bibnamefont {Tordeux}},\ }\bibfield
  {title} {\bibinfo {title} {Spontaneous wave formation in stochastic
  self-driven particle systems},\ }\href@noop {} {\bibfield  {journal}
  {\bibinfo  {journal} {SIAM Journal on Applied Mathematics}\ }\textbf
  {\bibinfo {volume} {81}},\ \bibinfo {pages} {853} (\bibinfo {year}
  {2021})}\BibitemShut {NoStop}%
\bibitem [{\citenamefont {Khound}\ \emph {et~al.}(2023)\citenamefont {Khound},
  \citenamefont {Will}, \citenamefont {Tordeux},\ and\ \citenamefont
  {Gronwald}}]{khound2021extending}%
  \BibitemOpen
  \bibfield  {author} {\bibinfo {author} {\bibfnamefont {P.}~\bibnamefont
  {Khound}}, \bibinfo {author} {\bibfnamefont {P.}~\bibnamefont {Will}},
  \bibinfo {author} {\bibfnamefont {A.}~\bibnamefont {Tordeux}},\ and\ \bibinfo
  {author} {\bibfnamefont {F.}~\bibnamefont {Gronwald}},\ }\bibfield  {title}
  {\bibinfo {title} {Extending the adaptive time gap car-following model to
  enhance local and string stability for adaptive cruise control systems},\
  }\href@noop {} {\bibfield  {journal} {\bibinfo  {journal} {Journal of
  Intelligent Transportation Systems}\ }\textbf {\bibinfo {volume} {27}},\
  \bibinfo {pages} {36} (\bibinfo {year} {2023})}\BibitemShut {NoStop}%
\bibitem [{\citenamefont {Ehrhardt}\ and\ \citenamefont
  {Tordeux}(2024)}]{ehrhardt2024stability}%
  \BibitemOpen
  \bibfield  {author} {\bibinfo {author} {\bibfnamefont {M.}~\bibnamefont
  {Ehrhardt}}\ and\ \bibinfo {author} {\bibfnamefont {A.}~\bibnamefont
  {Tordeux}},\ }\bibfield  {title} {\bibinfo {title} {Stability of
  heterogeneous linear and nonlinear car-following models},\ }\href@noop {}
  {\bibfield  {journal} {\bibinfo  {journal} {Franklin Open}\ }\textbf
  {\bibinfo {volume} {9}},\ \bibinfo {pages} {100181} (\bibinfo {year}
  {2024})}\BibitemShut {NoStop}%
\bibitem [{\citenamefont {Chen}\ \emph {et~al.}(2012)\citenamefont {Chen},
  \citenamefont {Laval}, \citenamefont {Ahn},\ and\ \citenamefont
  {Zheng}}]{chen2012microscopic}%
  \BibitemOpen
  \bibfield  {author} {\bibinfo {author} {\bibfnamefont {D.}~\bibnamefont
  {Chen}}, \bibinfo {author} {\bibfnamefont {J.~A.}\ \bibnamefont {Laval}},
  \bibinfo {author} {\bibfnamefont {S.}~\bibnamefont {Ahn}},\ and\ \bibinfo
  {author} {\bibfnamefont {Z.}~\bibnamefont {Zheng}},\ }\bibfield  {title}
  {\bibinfo {title} {Microscopic traffic hysteresis in traffic oscillations: A
  behavioral perspective},\ }\href@noop {} {\bibfield  {journal} {\bibinfo
  {journal} {Transportation Research Part B: Methodological}\ }\textbf
  {\bibinfo {volume} {46}},\ \bibinfo {pages} {1440} (\bibinfo {year}
  {2012})}\BibitemShut {NoStop}%
\bibitem [{\citenamefont {Orosz}\ \emph
  {et~al.}(2004{\natexlab{b}})\citenamefont {Orosz}, \citenamefont {Wilson},\
  and\ \citenamefont {Krauskopf}}]{orosz_GlobalBifurcationInvestigation_2004}%
  \BibitemOpen
  \bibfield  {author} {\bibinfo {author} {\bibfnamefont {G.}~\bibnamefont
  {Orosz}}, \bibinfo {author} {\bibfnamefont {R.}~\bibnamefont {Wilson}},\ and\
  \bibinfo {author} {\bibfnamefont {B.}~\bibnamefont {Krauskopf}},\ }\bibfield
  {title} {\bibinfo {title} {Global bifurcation investigation of an optimal
  velocity traffic model with driver reaction time},\ }\href
  {https://doi.org/10.1103/PhysRevE.70.026207} {\bibfield  {journal} {\bibinfo
  {journal} {Physical Review E}\ }\textbf {\bibinfo {volume} {70}},\ \bibinfo
  {pages} {026207} (\bibinfo {year} {2004}{\natexlab{b}})}\BibitemShut
  {NoStop}%
\bibitem [{Note1()}]{Note1}%
  \BibitemOpen
  \bibinfo {note} {The idea of introducing oscillations also pops up when one
  writes an amplitude equation in hydrodynamics \cite {morozov2005subcritical},
  but in that case it helps probe the nonlinear growth of arbitrary
  perturbations.}\BibitemShut {Stop}%
\bibitem [{\citenamefont {Kerner}\ and\ \citenamefont
  {Rehborn}(1997)}]{KernerR97}%
  \BibitemOpen
  \bibfield  {author} {\bibinfo {author} {\bibfnamefont {B.}~\bibnamefont
  {Kerner}}\ and\ \bibinfo {author} {\bibfnamefont {H.}~\bibnamefont
  {Rehborn}},\ }\bibfield  {title} {\bibinfo {title} {Experimental properties
  of phase transitions in traffic flow},\ }\href
  {https://doi.org/10.1103/PhysRevLett.79.4030} {\bibfield  {journal} {\bibinfo
   {journal} {Physical Review Letters}\ }\textbf {\bibinfo {volume} {79}},\
  \bibinfo {pages} {4030} (\bibinfo {year} {1997})}\BibitemShut {NoStop}%
\bibitem [{\citenamefont {Nagatani}(2002)}]{nagatani2002physics}%
  \BibitemOpen
  \bibfield  {author} {\bibinfo {author} {\bibfnamefont {T.}~\bibnamefont
  {Nagatani}},\ }\bibfield  {title} {\bibinfo {title} {The physics of traffic
  jams},\ }\href@noop {} {\bibfield  {journal} {\bibinfo  {journal} {Reports on
  Progress in Physics}\ }\textbf {\bibinfo {volume} {65}},\ \bibinfo {pages}
  {1331} (\bibinfo {year} {2002})}\BibitemShut {NoStop}%
\bibitem [{\citenamefont {Nagatani}(1998)}]{nagatani1998thermodynamic}%
  \BibitemOpen
  \bibfield  {author} {\bibinfo {author} {\bibfnamefont {T.}~\bibnamefont
  {Nagatani}},\ }\bibfield  {title} {\bibinfo {title} {Thermodynamic theory for
  the jamming transition in traffic flow},\ }\href@noop {} {\bibfield
  {journal} {\bibinfo  {journal} {Physical Review E}\ }\textbf {\bibinfo
  {volume} {58}},\ \bibinfo {pages} {4271} (\bibinfo {year}
  {1998})}\BibitemShut {NoStop}%
\bibitem [{\citenamefont {Jost}\ and\ \citenamefont
  {Nagel}(2005)}]{jost2005probabilistic}%
  \BibitemOpen
  \bibfield  {author} {\bibinfo {author} {\bibfnamefont {D.}~\bibnamefont
  {Jost}}\ and\ \bibinfo {author} {\bibfnamefont {K.}~\bibnamefont {Nagel}},\
  }\bibfield  {title} {\bibinfo {title} {Probabilistic traffic flow breakdown
  in stochastic car following models},\ }in\ \href
  {https://doi.org/https://doi.org/10.1007/3-540-28091-X_9} {\emph {\bibinfo
  {booktitle} {Traffic and Granular Flow '03'}}},\ \bibinfo {editor} {edited
  by\ \bibinfo {editor} {\bibfnamefont {S.}~\bibnamefont {Hoogendoorn}},
  \bibinfo {editor} {\bibfnamefont {S.}~\bibnamefont {Luding}}, \bibinfo
  {editor} {\bibfnamefont {P.}~\bibnamefont {Bovy}}, \bibinfo {editor}
  {\bibfnamefont {M.}~\bibnamefont {Schreckenberg}},\ and\ \bibinfo {editor}
  {\bibfnamefont {D.}~\bibnamefont {Wolf}}}\ (\bibinfo  {publisher}
  {Springer},\ \bibinfo {address} {Berlin, Heidelberg},\ \bibinfo {year}
  {2005})\ pp.\ \bibinfo {pages} {88--103}\BibitemShut {NoStop}%
\bibitem [{\citenamefont {Nagel}\ \emph {et~al.}(2003)\citenamefont {Nagel},
  \citenamefont {Wagner},\ and\ \citenamefont {Woesler}}]{NagelWW03}%
  \BibitemOpen
  \bibfield  {author} {\bibinfo {author} {\bibfnamefont {K.}~\bibnamefont
  {Nagel}}, \bibinfo {author} {\bibfnamefont {P.}~\bibnamefont {Wagner}},\ and\
  \bibinfo {author} {\bibfnamefont {R.}~\bibnamefont {Woesler}},\ }\bibfield
  {title} {\bibinfo {title} {Still flowing: Approaches to traffic flow and
  traffic jam modeling},\ }\href {https://doi.org/10.1287/opre.51.5.681.16755}
  {\bibfield  {journal} {\bibinfo  {journal} {Operations Research}\ }\textbf
  {\bibinfo {volume} {51}},\ \bibinfo {pages} {681} (\bibinfo {year}
  {2003})}\BibitemShut {NoStop}%
\bibitem [{\citenamefont {Helbing}\ \emph {et~al.}(2000)\citenamefont
  {Helbing}, \citenamefont {Farkas},\ and\ \citenamefont
  {Vicsek}}]{helbing2000freezing}%
  \BibitemOpen
  \bibfield  {author} {\bibinfo {author} {\bibfnamefont {D.}~\bibnamefont
  {Helbing}}, \bibinfo {author} {\bibfnamefont {I.~J.}\ \bibnamefont
  {Farkas}},\ and\ \bibinfo {author} {\bibfnamefont {T.}~\bibnamefont
  {Vicsek}},\ }\bibfield  {title} {\bibinfo {title} {Freezing by heating in a
  driven mesoscopic system},\ }\href@noop {} {\bibfield  {journal} {\bibinfo
  {journal} {Physical Review Letters}\ }\textbf {\bibinfo {volume} {84}},\
  \bibinfo {pages} {1240} (\bibinfo {year} {2000})}\BibitemShut {NoStop}%
\bibitem [{\citenamefont {Fruchart}\ \emph {et~al.}(2021)\citenamefont
  {Fruchart}, \citenamefont {Hanai}, \citenamefont {Littlewood},\ and\
  \citenamefont {Vitelli}}]{fruchart2021non}%
  \BibitemOpen
  \bibfield  {author} {\bibinfo {author} {\bibfnamefont {M.}~\bibnamefont
  {Fruchart}}, \bibinfo {author} {\bibfnamefont {R.}~\bibnamefont {Hanai}},
  \bibinfo {author} {\bibfnamefont {P.~B.}\ \bibnamefont {Littlewood}},\ and\
  \bibinfo {author} {\bibfnamefont {V.}~\bibnamefont {Vitelli}},\ }\bibfield
  {title} {\bibinfo {title} {Non-reciprocal phase transitions},\ }\href@noop {}
  {\bibfield  {journal} {\bibinfo  {journal} {Nature}\ }\textbf {\bibinfo
  {volume} {592}},\ \bibinfo {pages} {363} (\bibinfo {year}
  {2021})}\BibitemShut {NoStop}%
\bibitem [{\citenamefont {Ke-Ping}\ and\ \citenamefont
  {Zi-You}(2004{\natexlab{b}})}]{ke2004noise}%
  \BibitemOpen
  \bibfield  {author} {\bibinfo {author} {\bibfnamefont {L.}~\bibnamefont
  {Ke-Ping}}\ and\ \bibinfo {author} {\bibfnamefont {G.}~\bibnamefont
  {Zi-You}},\ }\bibfield  {title} {\bibinfo {title} {Noise-induced phase
  transition in traffic flow},\ }\href@noop {} {\bibfield  {journal} {\bibinfo
  {journal} {Communications in Theoretical Physics}\ }\textbf {\bibinfo
  {volume} {42}},\ \bibinfo {pages} {369} (\bibinfo {year}
  {2004}{\natexlab{b}})}\BibitemShut {NoStop}%
\bibitem [{\citenamefont {Krau{\ss}}\ \emph {et~al.}(1996)\citenamefont
  {Krau{\ss}}, \citenamefont {Wagner},\ and\ \citenamefont
  {Gawron}}]{krauss1996continuous}%
  \BibitemOpen
  \bibfield  {author} {\bibinfo {author} {\bibfnamefont {S.}~\bibnamefont
  {Krau{\ss}}}, \bibinfo {author} {\bibfnamefont {P.}~\bibnamefont {Wagner}},\
  and\ \bibinfo {author} {\bibfnamefont {C.}~\bibnamefont {Gawron}},\
  }\bibfield  {title} {\bibinfo {title} {{Continuous limit of the
  Nagel-Schreckenberg model}},\ }\href@noop {} {\bibfield  {journal} {\bibinfo
  {journal} {Physical Review E}\ }\textbf {\bibinfo {volume} {54}},\ \bibinfo
  {pages} {3707} (\bibinfo {year} {1996})}\BibitemShut {NoStop}%
\bibitem [{\citenamefont {Tordeux}\ and\ \citenamefont
  {Schadschneider}(2016)}]{Tordeux2016b}%
  \BibitemOpen
  \bibfield  {author} {\bibinfo {author} {\bibfnamefont {A.}~\bibnamefont
  {Tordeux}}\ and\ \bibinfo {author} {\bibfnamefont {A.}~\bibnamefont
  {Schadschneider}},\ }\bibfield  {title} {\bibinfo {title} {White and relaxed
  noises in optimal velocity models for pedestrian flow with stop-and-go
  waves},\ }\href@noop {} {\bibfield  {journal} {\bibinfo  {journal} {Journal
  of Physics A: Mathematical and Theoretical}\ }\textbf {\bibinfo {volume}
  {18}},\ \bibinfo {pages} {185101} (\bibinfo {year} {2016})}\BibitemShut
  {NoStop}%
\bibitem [{\citenamefont {Eilhardt}\ and\ \citenamefont
  {Schadschneider}(2015)}]{EilhardtS15}%
  \BibitemOpen
  \bibfield  {author} {\bibinfo {author} {\bibfnamefont {C.}~\bibnamefont
  {Eilhardt}}\ and\ \bibinfo {author} {\bibfnamefont {A.}~\bibnamefont
  {Schadschneider}},\ }\bibfield  {title} {\bibinfo {title} {Stochastic headway
  dependent velocity model and phase separation in pedestrian dynamics},\ }in\
  \href@noop {} {\emph {\bibinfo {booktitle} {Traffic and {{Granular Flow}}
  2013}}},\ \bibinfo {editor} {edited by\ \bibinfo {editor} {\bibfnamefont
  {M.}~\bibnamefont {Chraibi}}, \bibinfo {editor} {\bibfnamefont
  {M.}~\bibnamefont {Boltes}}, \bibinfo {editor} {\bibfnamefont
  {A.}~\bibnamefont {Schadschneider}},\ and\ \bibinfo {editor} {\bibfnamefont
  {A.}~\bibnamefont {Seyfried}}}\ (\bibinfo  {publisher} {Springer},\ \bibinfo
  {year} {2015})\ p.\ \bibinfo {pages} {382}\BibitemShut {NoStop}%
\bibitem [{\citenamefont {Morozov}\ and\ \citenamefont {van
  Saarloos}(2005)}]{morozov2005subcritical}%
  \BibitemOpen
  \bibfield  {author} {\bibinfo {author} {\bibfnamefont {A.~N.}\ \bibnamefont
  {Morozov}}\ and\ \bibinfo {author} {\bibfnamefont {W.}~\bibnamefont {van
  Saarloos}},\ }\bibfield  {title} {\bibinfo {title} {Subcritical
  finite-amplitude solutions for plane couette flow of viscoelastic fluids},\
  }\href@noop {} {\bibfield  {journal} {\bibinfo  {journal} {Physical Review
  Letters}\ }\textbf {\bibinfo {volume} {95}},\ \bibinfo {pages} {024501}
  (\bibinfo {year} {2005})}\BibitemShut {NoStop}%
\bibitem [{\citenamefont {Ngoduy}(2015)}]{ngoduy2015effect}%
  \BibitemOpen
  \bibfield  {author} {\bibinfo {author} {\bibfnamefont {D.}~\bibnamefont
  {Ngoduy}},\ }\bibfield  {title} {\bibinfo {title} {Effect of the
  car-following combinations on the instability of heterogeneous traffic
  flow},\ }\href@noop {} {\bibfield  {journal} {\bibinfo  {journal}
  {Transportmetrica B: Transport Dynamics}\ }\textbf {\bibinfo {volume} {3}},\
  \bibinfo {pages} {44} (\bibinfo {year} {2015})}\BibitemShut {Stop}%
\end{thebibliography}
\end{document}